\documentclass{sigchi}

\CopyrightYear{2017} 
\setcopyright{acmlicensed} 
\conferenceinfo{UIST 2017,}{October 22--25, 2017, Quebec City, QC, Canada}
\isbn{978-1-4503-4981-9/17/10}\acmPrice{\$15.00}
\doi{https://doi.org/10.1145/3126594.3126635}

\usepackage{balance}       
\usepackage{graphics}      
\usepackage[T1]{fontenc}   
\usepackage{txfonts}
\usepackage[pdflang={en-US},pdftex]{hyperref}
\usepackage{color}
\usepackage{booktabs}
\usepackage{textcomp}
\usepackage{wrapfig}

\usepackage{soul}

\usepackage{ccicons}          

\usepackage{todonotes}

\teaser{
\vspace{-3mm}
\centering
  \includegraphics[width=0.95\linewidth]{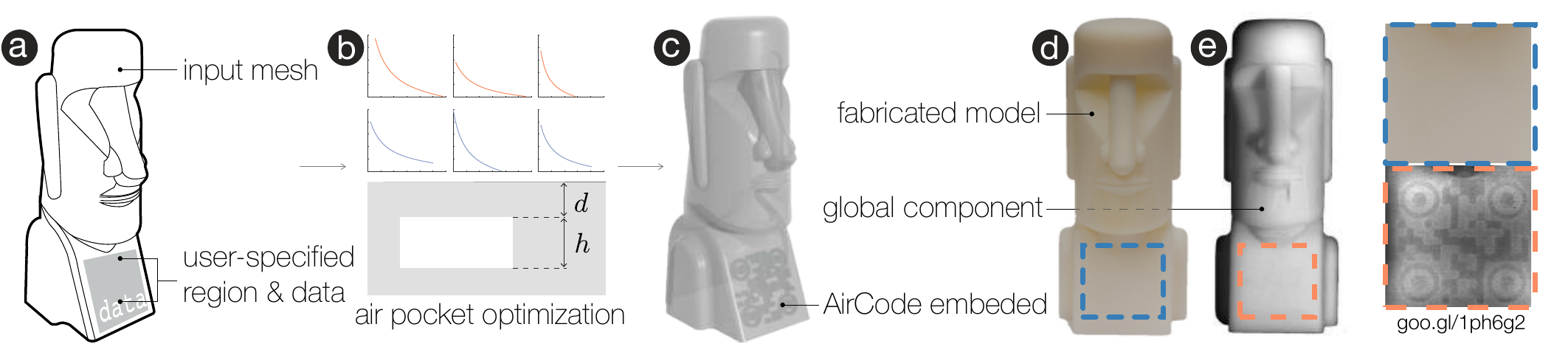}
\vspace{-1mm}
  \caption{
  (a) AirCode tagging tool takes as inputs a mesh, a user-specified region, and embedded data.
  (b) It first determines the air pocket parameters, including the depth $d$ and thickness $h$, for a fabrication material.
  (c) An AirCode tag is embedded inside the object, without changing its geometry or appearance.
  (d) The fabricated tag is invisible under environmental lighting.
  (e) Using our imaging system that separates out the 
  global scattering effects, 
  the user detects the embedded tag and retrieve the data.
  ``\href{https://www.thingiverse.com/thing:2422543}{Moai}'' by gravityisweak /
  \href{https://creativecommons.org/licenses/by/3.0/}{CC BY 3.0} / modified from original.
   \label{fig:teaser}
   }
\vspace{-2mm}
}

\newcommand{\papername}[0]{{AirCode}}
\def\plaintitle{AirCode: Unobtrusive Physical Tags for Digital Fabrication}

\def\emptyauthor{}
\def\plainkeywords{digital fabrication; 3D printing; unobtrusive tags; air pockets; sensing}

\makeatletter
\def\url@leostyle{%
  \@ifundefined{selectfont}{
    \def\UrlFont{\sf}
  }{
    \def\UrlFont{\small\bf\ttfamily}
  }}
\makeatother
\def\pprw{8.5in}
\def\pprh{11in}

\definecolor{linkColor}{RGB}{6,125,233}
\hypersetup{%
  pdftitle={\plaintitle},
  pdfauthor={\emptyauthor},
  pdfkeywords={\plainkeywords},
  pdfdisplaydoctitle=true, 
  bookmarksnumbered,
  pdfstartview={FitH},
  colorlinks,
  citecolor=black,
  filecolor=black,
  linkcolor=black,
  urlcolor=linkColor,
  breaklinks=true,
  hypertexnames=false
}

\usepackage{amssymb}
\usepackage{bm}
\usepackage{mathtools}

\usepackage{algorithm}
\usepackage{algpseudocode}
\algtext*{EndWhile}
\algtext*{EndIf}
\algtext*{EndProcedure}

\newcommand{\sLNM}{\begin{linenomath}}
\newcommand{\tLNM}{\end{linenomath}}

\newcommand{\figref}[1]{Figure~\ref{#1}}

\newcommand{\eqnref}[1]{Eq.~\eqref{#1}}

\newcommand{\bx}[0]{\bm{x}}
\newcommand{\bomg}[0]{\bm{\omega}}
\newcommand{\dd}[0]{\textrm{d}}
\newcommand{\calT}[0]{\mathcal{T}}
\newcommand{\calR}[0]{\mathcal{R}}

\newcommand{\ssa}[0]{\mathsf{a}}
\newcommand{\ssc}[0]{\mathsf{c}}

\begin{document}

\title{\plaintitle}

\numberofauthors{1}
\author{%
  \alignauthor{Dingzeyu Li ~~~~~~~ Avinash S. Nair ~~~~~~~ Shree K. Nayar ~~~~~~~ Changxi Zheng\\
    \affaddr{Columbia University, New York, NY}\\
    \email{\{dli,nayar,cxz\}@cs.columbia.edu}, asn2129@columbia.edu}\\
    }

\maketitle

\begin{abstract}

We present \emph{AirCode}, a technique that allows the user 
to tag physically fabricated objects with given information.
An AirCode tag consists of a group of carefully designed air pockets placed 
beneath the object surface.
These air pockets are easily produced during the fabrication process of the object, without 
any additional material or postprocessing.
Meanwhile, the air pockets affect only the scattering light transport under the surface,
and thus are hard to notice to our naked eyes.
But, by using a computational imaging method, the tags become detectable.
We present a tool that automates the design of air pockets for the user to encode information.
AirCode system also allows the user to retrieve the information from captured images
via a robust decoding algorithm.
We demonstrate our tagging technique with applications
for metadata embedding, robotic grasping, as well as conveying object affordances. 

\end{abstract}


\category{H.5.m.}{Information Interfaces and Presentation
 (e.g. HCI)}{Miscellaneous} 

\newcommand{\rev}[1]{{{ #1}}}

\section{Introduction}\label{sec:intro}

Whether we board airplanes, borrow books from a library, or line up to check
out at grocery stores, one common minutia we benefit from is the optical tag,
a machine-readable, black-and-white pattern printed on a surface to contain
information about the item on which it is printed. Today, optical tags have
become a technological staple of everyday life, establishing ``hyperlinks''
between physical surfaces and digital information.

In this paper, we extend the idea of hyperlinks and propose \emph{\papername}, an unobtrusive tagging system for 3D printed objects.
3D printing has the unprecedented ability to create customized, one-off parts,
necessitating tags that carry individualized information.
For instance, when fabricating many similarly shaped components that are assembled together, 
 it would be beneficial to tag each component to facilitate correct assembly.
Physical tags also establish a link between physically manufactured objects and digital 
computing systems: a robot can better recognize a 3D printed object and its
poses for manipulation, by reading tags attached to the object.

In developing a practical tagging system for 3D printing, several desiderata are of importance.
\textbf{(i)} Tags need to be embedded during the 3D printing process, not as a separate post-processing step.
This is because post-processing not only introduces extra cost but 
requires one to distinguish individual objects in the first place---a step that by itself benefits from tags. 
\textbf{(ii)} Tags need to be printable with existing 3D printers.
Ideally, even a single-material printer should be able to tag its fabrication.
\textbf{(iii)} Tags need to be unobtrusive with respect to the shapes and appearance of 3D printed objects.

To our knowledge, none of the existing solutions satisfies these requirements.
For example, traditional optical tags fail with respect to (i) and (iii),
and Radio Frequency Identification (RFID) tags break (i) and (ii).

{AirCode} satisfies all above requirements.
Our key idea is simply placing thin air pockets under the surface of 3D printed objects.
Without requiring any additional material or post processing, 
air pockets can be easily produced by most 3D printers.
Meanwhile, air, drastically different from 3D printing materials in terms of optical properties, 
changes how light is scattered after penetrating the material surface.

Most plastic 3D printing materials, even those considered opaque, scatter light,
while the amount of light penetrating and scattered is often weak; most of the
light is directly reflected at the surface. Consequently, the effects of air pockets 
on object appearance can be made imperceptible to our naked eyes.
But the user can separate out the subsurface scattered light through a 
computational imaging method 
that requires only a conventional camera and projector,
and in turn amplify the light transport effects of air pockets.
We demonstrate that by carefully designing the subsurface air pockets, one can
conceal information in imperceptible yet machine-readable tags.

\begin{figure}[t]
  \vspace{-0.5mm}
  \centering
  \includegraphics[width=0.92\columnwidth]{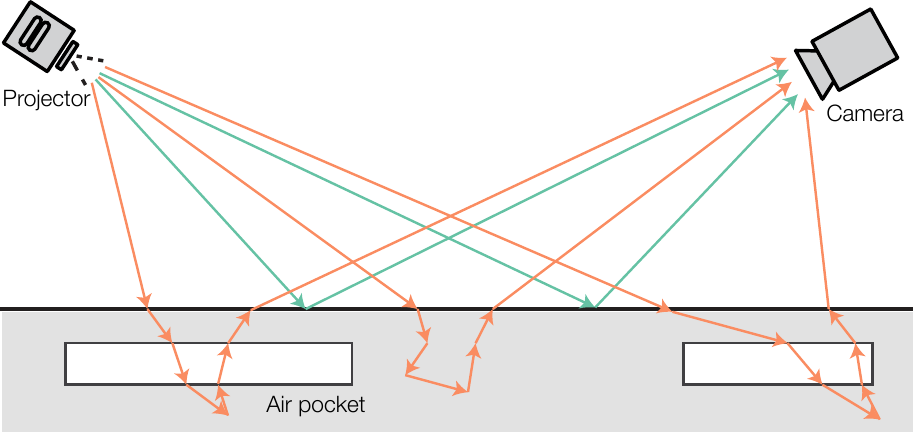}
  \caption{
  Key observation: Most plastic 3D printing materials
   exhibit strong subsurface scattering.
  Light rays (green) that are reflected by the surface represent the direct component;
  rays (orange) that enter the surface and are scattered within before leaving the surface result in the global component. 
  A structured change in the material that lies beneath the surface only affects the global component of a captured image.
  \label{fig:illu}}
  \vspace{-4mm}
\end{figure}

We present a design tool that determines the shapes and positions of 
subsurface air pockets to encode user-specified information.
Our system also enables the user to separate the global illumination
light transport from the direct illumination,
using computational imaging~\cite{NayarKGR06}. 
The direct component accounts for light rays reflected by the object
surface and thus is unaffected by subsurface air pockets. The global
component is dominated by light rays that are scattered after penetrating the surface (\figref{fig:illu}).
It is affected by the air pockets and thus conveys the embedded information.
Meanwhile, it is unaffected by
direct illumination effects such as specular highlights which often frustrate
machine vision systems. As a result, our method of reading subsurface tags is
robust to variation in object pose and camera angle.

\section{Related Work}

The emergence of rapid fabrication tools allows users to 
prototype personal objects for fabrication~\cite{gershenfeld2008fab}.
Recently, HCI researchers have created various design tools to facilitate 
the design process~\cite{DBLP:conf/chi/CW16,DBLP:conf/uist/McCraeUS14,DBLP:conf/uist/TeibrichMGKNB15,DBLP:conf/chi/VasilevitskyZ16}. 
For example, On-the-fly Printing enables incremental printing during 
the modeling stage~\cite{DBLP:conf/chi/PengWMG16}; ChronoFab, a 3D modeling tool, allows
creating motion sculptures~\cite{DBLP:conf/chi/KaziGMSF16}.
The size of a personal fabricated object can range from hair fibers~\cite{DBLP:conf/uist/LaputCH15},
palm-size pieces~\cite{DBLP:conf/chi/MuellerMGFB14,DBLP:conf/uist/IonFWKALLCB16},
to room-size objects~\cite{DBLP:conf/uist/AgrawalUKFCMB15}.
To facilitate interaction with these customized objects, unobtrusive physical
tags are desired to link physical objects with digital systems.

In order to apply a tag to a physical object, perhaps among the first choices is the
conventional optical barcode, such as a 1D linear barcode and a 2D QR code.
Optical barcodes have been used in applications ranging from
augmented reality marking, robot-human interaction, 
to context-aware gesture interaction~\cite{DBLP:conf/cvpr/Fiala05,DBLP:conf/chi/ZhaoNII09,DBLP:conf/uist/ChanCHLC15}.
Despite their straightforward use on digital displays or printed materials, they suffer from some limitations.
First, barcodes are obtrusive, sometimes even distracting, since
they always occupy a surface region and are visible to our naked eyes.
Second, the decoding process often requires a relatively clean and sharp image. 
But when we embed the barcode beneath a surface, the imaging results become
blurry, noisy, and of low contrast (\figref{fig:qrcorner}).
Our AirCode system is inspired by existing barcodes but 
tailored to enable robust decoding.

RFID is also commonly used for tagging but typically requires a postprocessing step
to install RFID circuits inside an object~\cite{SpielbergSHMM16}.  
In contrast, our proposed method requires no postprocessing.  This is desirable
if one needs a fully automatic pipeline wherein a robotic system can manipulate
the object immediately after its manufacturing.\rev{Recent work on 3D printing
of electronics~\cite{espalin20143d} is promising for making objects with embedded RFIDs.
But this technology is not yet accessible to most users, 
and the fabrication is also more costly in comparison to our method relying on only commodity 3D printers. 
Essentially an optical code, AirCode complements RFID tags.
For example, it is easy to estimate object orientation using optical codes
while not straightforward for RFIDs.}

\rev{Another approach is printing with invisible inks that can be revealed 
under ultraviolet light~\cite{hu2013invisible}.
An additional process (after the fabrication) is needed to color the object
with UV-visible inks. 
Moreover, UV ink can fade under direct exposure to
lighting or wear off after prolonged surface interaction with users and other
objects, whereas air pockets used in AirCode tags are well shielded under the object surface.}

Recent work has explored other tagging mechanisms such as 
encoding in the time sequence of audio signals or the acoustic 
frequency spectrum~\cite{DBLP:conf/uist/HarrisonXH12,LaputBHH15,Li:2016:acoustic_voxels,DBLP:conf/chi/SavageHHGML15}.
For instance, Acoustic Barcodes~\cite{DBLP:conf/uist/HarrisonXH12} encode
binary IDs in structured patterns of physical notches on an object surface, and
the IDs are machine readable via analyzing the sound produced by swiping the
notches.  This type of methods requires changing the shape of the object when
applying tags.  In addition, physical contacts are required to read the tags,
whereas we are able to detect an \papername~tag through a camera system,
without touching or knowing the exact location of the object.  

%

Sharing a similar goal to our approach, Willis et al.~\shortcite{WillisW13} 
used Terahertz (THz) imaging devices to scan internal structures of 3D printed objects.
While demonstrating promising results, 
these methods require expensive imaging equipment
and are limited by a relatively low spatial imaging resolution
 (e.g., 30$\times$30 as reported in \cite{WillisW13}).
In contrast, AirCode tags can be captured by a conventional, low-cost
camera system and produce high-resolution images of the tags.

With the emergence of advanced 3D printers, subsurface scattering has been 
recently exploited for appearance fabrication~\cite{HasanFMPR10,PapasRJBJMMG13}.
As a pioneer work in realistic rendering, Jensen et al.~\cite{jensen2001practical}
modeled subsurface scattering using dipole approximation to the 
bidirectional scattering surface reflectance distribution functions (BSSRDF).
Based on these approximations, to fabricate a desired translucent appearance,
Ha\u{s}an et al.~\shortcite{HasanFMPR10} composite layered materials to 
obtain user-specified BSSRDFs in 3D printed objects, 
In comparison, our work is not meant to physically reproduce specific material appearance.
Instead, we aim to \emph{preserve} the appearance of objects while embedding information in them.

\section{Method Overview} 

Our framework powering AirCode exploits subsurface light scattering to 
design tags that are imperceptible but machine-readable.
Our system consists of three major steps.

\paragraph{Preprocessing: Determining Air Pocket Parameters}
Provided a fabrication material, our system first determines the geometric parameters of
air pockets---the parameters that describe the size and depth
of subsurface air pockets (\figref{fig:teaser}-b) such that the air pockets are
invisible to the human eye, and meanwhile produce sufficiently clear
features on the global-component image for reliable tag reading.
To determine these parameters, we measure the material's subsurface scattering
properties and in turn analyze the influence of air pockets on the material's surface 
appearance. This step is a one-time process for a given fabrication material.


\paragraph{AirCode Design}
When the user specifies a piece of information (represented as a bit string) and a 3D
model for fabrication, our system generates a layout of air pockets placed beneath the surface
of the 3D model without changing its surface shape.
Each air pocket in this layout is constructed with the parameters estimated in the preprocessing
step. The air pocket layout serves two purposes: 
(i) it enables the reading algorithm to robustly locate AirCode tags on a global-component image,
and (ii) it embeds the given information.
The output of this step is a 3D model with air pockets embedded, ready for physical fabrication.



\paragraph{AirCode Reading}
Our method to read AirCode tags that are embedded in a physical object is based on a 
computational imaging 
technique~\cite{NayarKGR06}. 
 The imaging method produces a direct and
a global component image, of which the latter conveys the influence of the
subsurface air pockets. 
Because of subsurface scattering, the global component image is blurry and of
low contrast. And 3D printing artifacts (such as the printhead motion patterns)
further introduce image noise.  Our system locates the tags on the
global-component image using a multi-scale elliptical detector and then
retrieve embedded information using an SVM classifier trained on the fly.


\section{Example Applications}
AirCode provides an unobtrusive way
of embedding user-specified information in physical objects, 
and the tag is produced in the process of fabrication, without any postprocessing.
Not only is AirCode directly applicable in rapid prototyping (such as 3D printing)
but also in many mass-produced products.
Here we describe a few applications enabled by AirCode tags.
We also refer to the supplemental video for the demonstration.


\subsection{Embedding Metadata in Physical Objects}

Many digital productions carry \emph{metadata}, a piece of information that
is not directly perceptible but can be retrieved to provide additional 
resources and digital identification~\cite{greenberg2005understanding}.
Perhaps the most well-known are the metadata embedded in photographs, providing information
such as capture date, exposure time, focal length, GPS location, and copyright.

AirCode provides a means of embedding similar metadata but for physical
objects.  It can serve as a metadata holder to provide additional
information, resources, copyright, and digital identification for 3D printed
objects. 
For example, after designing an artistic statue, the artist can embed
a link to a webpage about the background of this statue or the artist's personal
website or copyright claim in the statue before fabricating it.  
Later, when the client receives the statue, by retrieving the embedded AirCode tags,
the client can learn more information and resources about the statue and the artist.


We demonstrate the use of \papername~with a Moai statue, as shown in \figref{fig:teaser}.
The user specifies a region to embed in the input model a link to the statue's webpage. 
The statue with this AirCode tag is then 3D printed. Since the tags are
embedded beneath the surface of the statue, they do not alter the geometry or
the appearance of the statue. However, by using our global-component imaging
system, the embedded tag can be detected and the webpage is retrieved.

%
%


\begin{figure}[t]
  \centering
  \includegraphics[width=0.95\columnwidth]{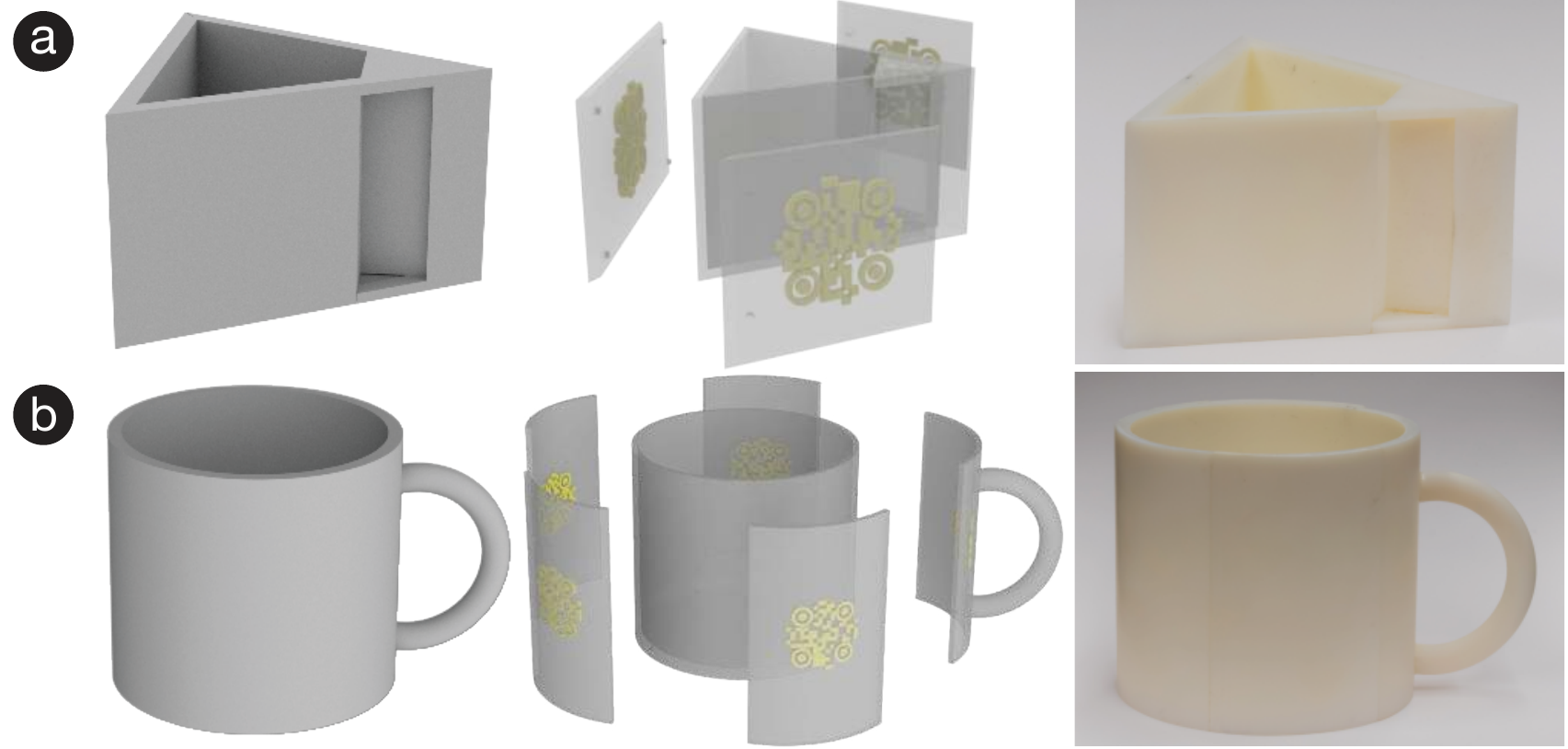}
  \caption{
      Multiple \papername~tags are embedded in a triangular drawer (top) and a mug (bottom).
      These tags are unique from each other.
      As long as one of the tags can be viewed by the camera system from a view angle,
      the object can be recognized and its pose can be estimated.
  \label{fig:show_drawer}}
  \vspace{-3mm}
\end{figure}

\begin{figure}[t]
	\centering
	\includegraphics[trim={0 0 15.6cm 0},clip,height=3cm]{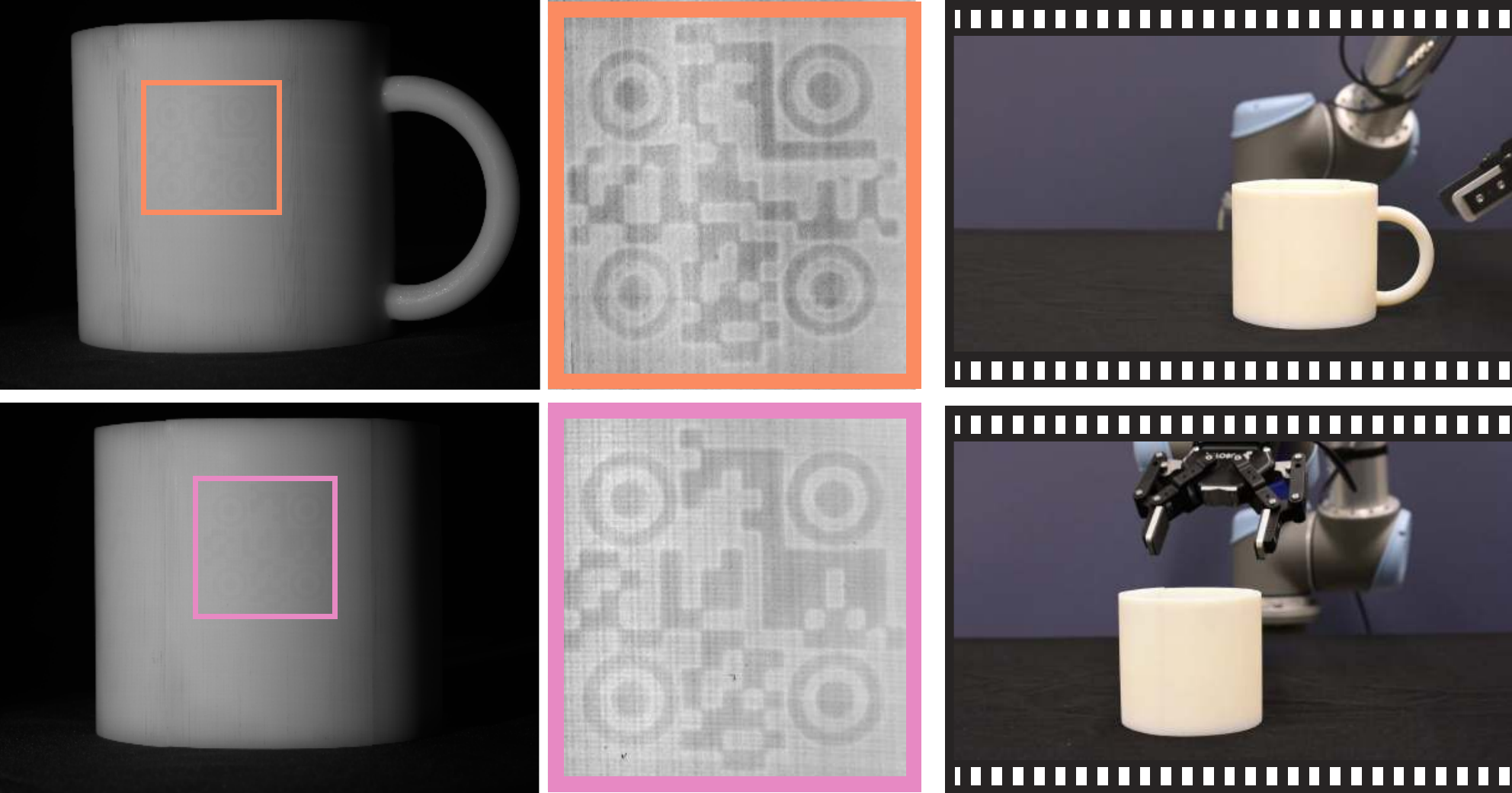}
	\hspace{-1.5mm}
	\includegraphics[trim={28.1cm 0 0 0},clip,height=3cm]{figs/figure_robot_mug.pdf}
	\caption{ 
{\bf Robotic grasping with occlusion.} We have demonstrated the use of our subsurface
codes for recognizing objects and determining their pose from the affine
deformation of the detected code in the image. This enables a robot to not only
identify the object but also plan the best grasping strategy needed to pick it
up. Note that since multiple codes are embedded in the object, the identity and
pose of the object is determined irrespective of its position and
orientation.
            \label{fig:robot_mug}}
        \vspace{-4mm}
\end{figure}

\subsection{Robotic Grasping}\label{sec:robot}
\papername~tags also help robotic manipulators interact with man-made objects. 
In robotic grasping tasks, a challenging problem is to recognize an object, estimate its pose, and 
decide where to grasp. Most robotic systems rely on the image camera and/or depth sensor to
infer the shape and pose of an object and plan grasping motion.
However, if a crucial region of an object (e.g., the handle of a mug) is occluded from the sensor,
it is very hard, if not impossible, for the robot to identify grasping points that are in the occluded region.

If an object embeds \papername~tags, the camera system of a robotic manipulator
can recognize the object and retrieve its complete 3D model by reading the
tags. More remarkably, as we will present in the AirCode Design and Reading
section, locating the \papername~tags further allows the system to estimate the
object pose (i.e., position and orientation) with respect to the camera.
Knowing the 3D model and the pose of an object, the robot has complete
information to plan the grasping.
We also note that since \papername~tags are generated in the process of object
fabrication, the robotic system can identify and manipulate the object
immediately after its fabrication, without any post processing to add optical
barcodes or RFIDs.  This is highly desirable for automated production pipelines
(such as those for automated assembly).



We demonstrate with two objects (see \figref{fig:show_drawer}). 
In the triangular drawer (\figref{fig:show_drawer}-{a}), 
we embed three different \papername~tags on each side of the drawer. 
In the mug (\figref{fig:show_drawer}-{b}), we embed six tags under the curved surface of the mug.
The tags are made unique from each other. As long as one of the tags is captured and read
by the imaging system, the robotic manipulator can identify the object 
and estimate its pose.


\papername~tags enable the robotic manipulator to sidestep the vision-based recognition problem
and directly identify the object. Consequently, the robot manipulator can grasp a handle that is completely 
occluded from the camera (see \figref{fig:robot_mug} and supplemental video). 
This is a particularly challenging case for vision-based
grasping methods because from a directly captured image, it is hard to infer
the parts that are occluded from the camera.

\begin{figure}[b]
  \vspace{-4mm}
  \centering
  \includegraphics[width=0.98\columnwidth]{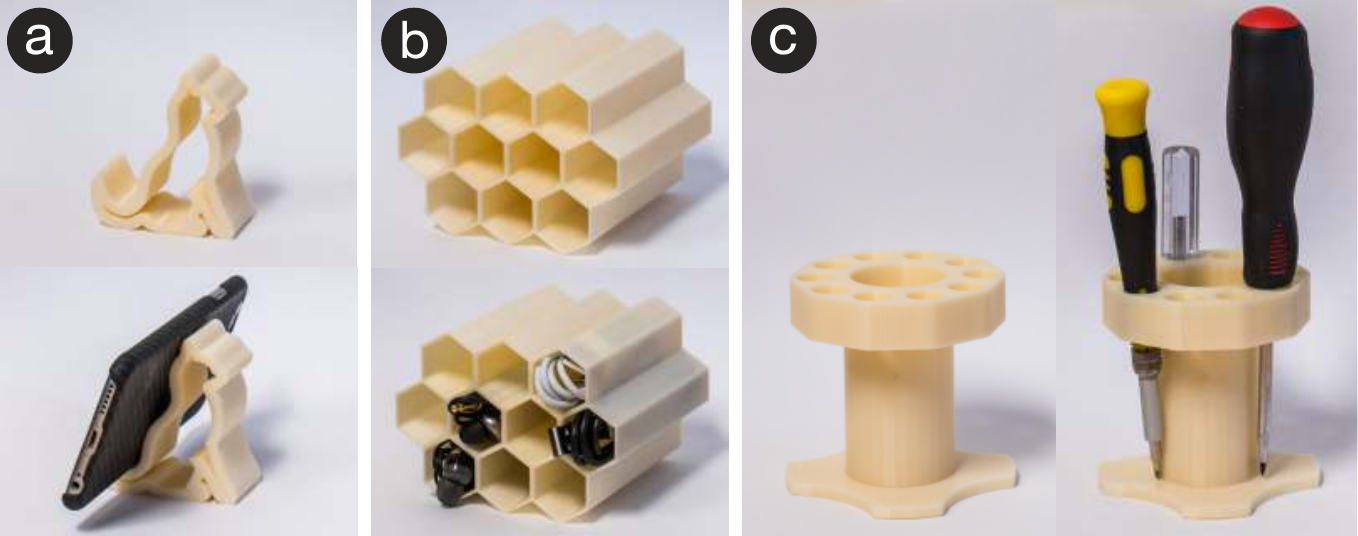}
  \caption{
      Three artistically designed objects for supporting an iPhone, 
      holding cables, and storing mechanical tools respectively are shown here.
      While aesthetically attractive, the affordances of these designs might not 
      be intuitively interpretable for an ordinary user.
      Embedding AirCode tags in these objects can help convey the affordances.
  ``\href{https://www.thingiverse.com/thing:702912}{Tool Carousel}'' by mbeyerle116 and
  ``\href{https://www.thingiverse.com/thing:1012788}{Kitty Phone Holder}'' by  Tinyeyes / 
  \href{https://creativecommons.org/licenses/by-sa/3.0/}{CC BY-SA 3.0}; 
  ``\href{https://www.thingiverse.com/thing:961176	}{Cable management Hive}'' by Filar3D /
  \href{https://creativecommons.org/licenses/by-nc/3.0/}{CC BY-NC 3.0}.
  \label{fig:affordance}}
\end{figure}

\subsection{Conveying Object Affordance}
Object affordance is ``the particular ways in which an actor, or set of
actors, perceives and uses an object''~\cite{gibson2014ecological,DBLP:conf/chi/LopesJB15}.  For
example, the handles on a teapot offer an obvious affordance for holding.
Some objects can be interpreted to afford different uses, while many customized objects
may not have an easily interpretable affordance. For example, \figref{fig:affordance} 
shows three objects from \textsf{Thingiverse}, a 3D model sharing website.
These models are designed with various intended functional purposes while offering unique 
aesthetics. However, the artistically designed shapes can conceal their 
affordances---for instance, the cat model (\figref{fig:affordance}-a) has a carefully designed distribution of mass in order 
to hold an iPhone stably,\rev{but this intended use can be unintuitive for the user to interpret.}

\papername~tags enable an unobtrusive way to embed information about an object's affordances
in the object itself during the design process. For example, the designer can embed 
 a link to a webpage that illustrates the  object's use without
sacrificing the appearance or the artistically designed shape of the object.  Then
the user can extract the link and understand its affordance.  Furthermore, the
embedded tags also allow a robotic system to know how to precisely manipulate
an object.  As customized 3D models become available
online, we envision that \papername~tags can help 
communicate their origins and uses more seamlessly.


%
%
%



\subsection{Extension: Paper Watermarking}
The idea of using air pockets to alter the global component of an image and thereby 
embed information can be extended\rev{beyond 3D printed objects or plastic materials.
In general, as long as the materials are not fully opaque, it is possible to exploit
subsurface light transport for tagging.}
As a demonstration of extending \papername~tags to other materials,
here we embed watermarks in a paper
by stacking a few thinner papers together. We carve a pattern on one paper and
sandwich it in other papers, and then stick all thin papers together.


\begin{figure}[t]
	\centering
	\includegraphics[width=1.0\columnwidth]{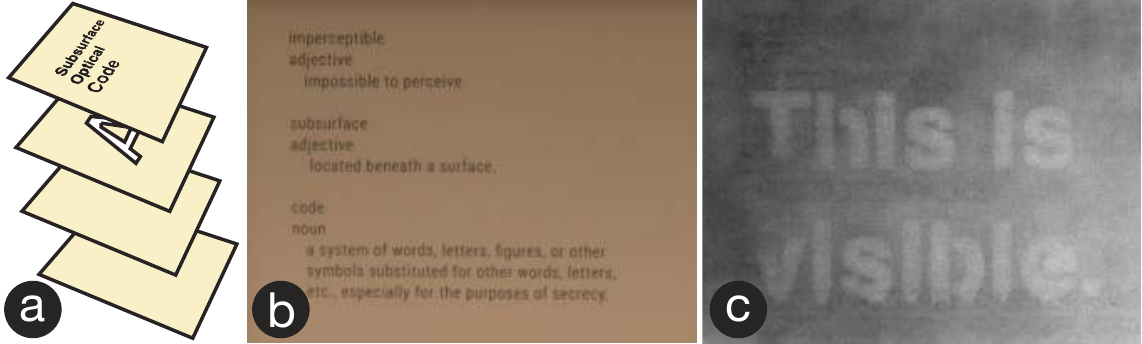}
        \vspace{-3mm}
	\caption{ {Paper Watermarking.}
AirCode tags can also be embedded within sheets of paper. {\bf (a)} A sheet
of paper is constructed from multiple thinner sheets where some of the sheets
have cut-outs that serve as air pockets. {\bf (b)} A conventional image of a sheet
with text printed on it. {\bf (c)} The global-component image recovered by the
imaging system reveals a hidden message. See text for details on how the
global-component image is processed to remove the effects of the printed seen
text in (b).
	}
	\label{fig:TPS}
        \vspace{-2mm}
\end{figure}

Particularly, the  paper watermark consists of four layers of thinner papers.  The top
layer is printed with regular visible text or data. The second layer is carved
with a stencil to hold a hidden message. The last two layers are blank
papers.  As shown in \figref{fig:TPS} and the video, the hidden message is
invisible under normal lighting conditions but can be detected using our
imaging method.  In this case, the surface texture (i.e., the printed text)
will also affect the global component image. We exploit the direct component
image which includes only the surface texture to create a mask for the printed
text.  With this mask, we can remove the printed text in the global component
image using a PatchMatch-based inpainting algorithm (e.g., featured in Adobe Photoshop).  We
envision that in the future this technique can be used as unobtrusive codes or
anti-forgery tags on product packages and books.


\section{AirCode Design and Reading}\label{sec:code}
\begin{figure}[t]
  \vspace{-1mm}
  \centering
  \includegraphics[width=0.99\columnwidth]{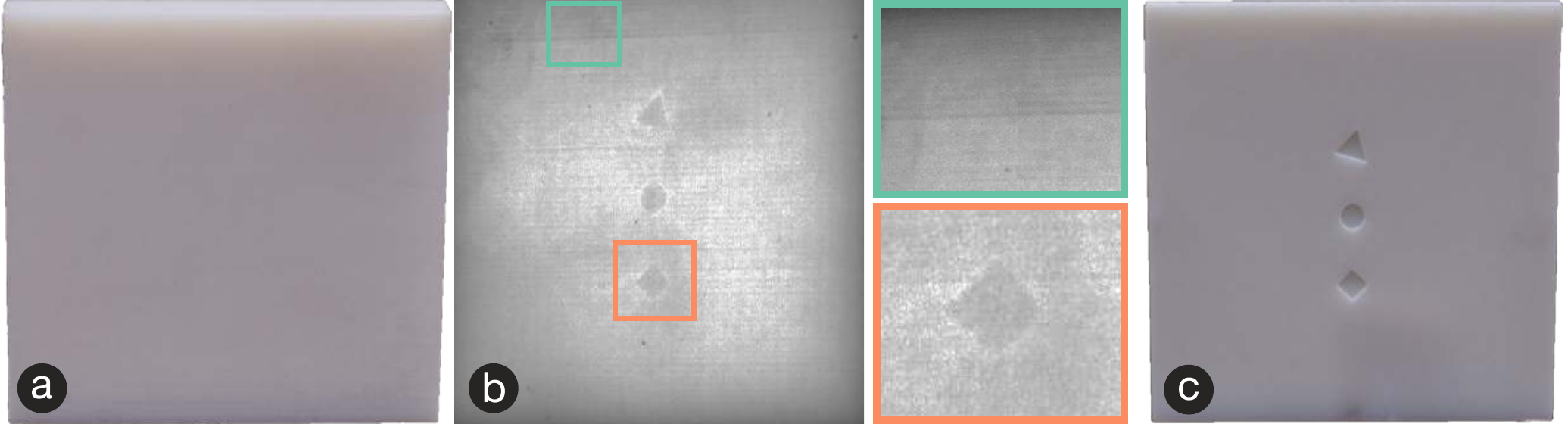}
  \caption{ Challenges in global-component imaging:
      {\bf (a)} A printed object with three air pockets within it. 
      {\bf (b)} The global-component image measured by the imaging system. 
      {\bf (c)} A cross-sectional view of the object which reveals the actual shapes of the
      air pockets.
  \label{fig:challenge}}
  \vspace{-1mm}
\end{figure}

We now present the core algorithmic components of our system, the AirCode
\emph{generation} and \emph{reading}. The basic element of an AirCode tag is an
air pocket placed beneath the object surface. We will present an analysis to 
estimate the geometric parameters of air pockets in Determining Air Pocket Parameters.
In this section, we focus on the layout of air pockets:
we first present a method to generate a subsurface air pocket structure that 
encodes user-provided information; we then describe our method that retrieves 
the embedded information from the global-component image of an object:




Our \papername~generation algorithm works in tandem with the imaging method that separates
the global illumination of light transport from the directional illumination.
We will describe the imaging details in the next section, but note here
the challenges arising from the global-component imaging, to rationalize 
algorithmic choices in our tag design:

\figref{fig:challenge} shows the cross section of a 3D printed model
with three air pockets placed under the surface and its global-component image. 
While the three air pockets are discernible from the global-component image, it is contaminated 
by the patterns of 3D printing filaments.
The air pocket shapes are blurred due to subsurface scattering 
and the intensity across the image is uneven.
With this observation, we seek a tag generation algorithm robust to these imaging artifacts.

\subsection{Encoding}\label{sec:encode}
To encode with an air pocket structure, we draw an analogy to the 
most popular two-dimensional barcode, the QR code.
The design of QR code has two crucial components:
(i) the concentric squares at the top-left, top-right, and bottom-left corners
and (ii) black-and-white cells each representing a 0/1 bit.
We refer the former as \emph{markers} and the latter as \emph{bits}.
The markers facilitate location of the QR code in a captured image. Their positions establish
a grid where bits are located, and the bits carry specific information.

\begin{figure}[b]
  \centering
  \vspace{-1.4mm}
  \includegraphics[width=0.99\columnwidth]{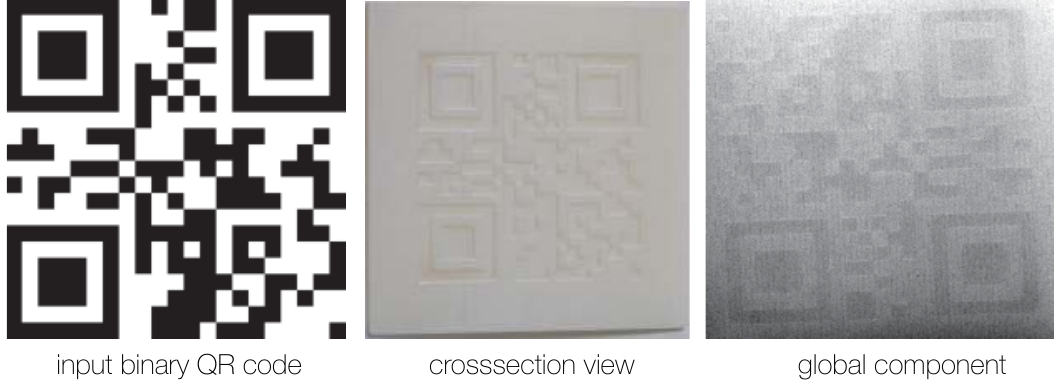}
  \caption{
      We experiment a naive placement of a QR code pattern (left) beneath an object surface.
      The cross section of a printed piece is shown in the middle. But the QR code pattern on the 
      global-component image (right) is too noisy to be decoded.
  \label{fig:qrcorner}}
  \vspace{-1.4mm}
\end{figure}

\begin{figure}[t]
  \centering
  \includegraphics[width=0.9\columnwidth]{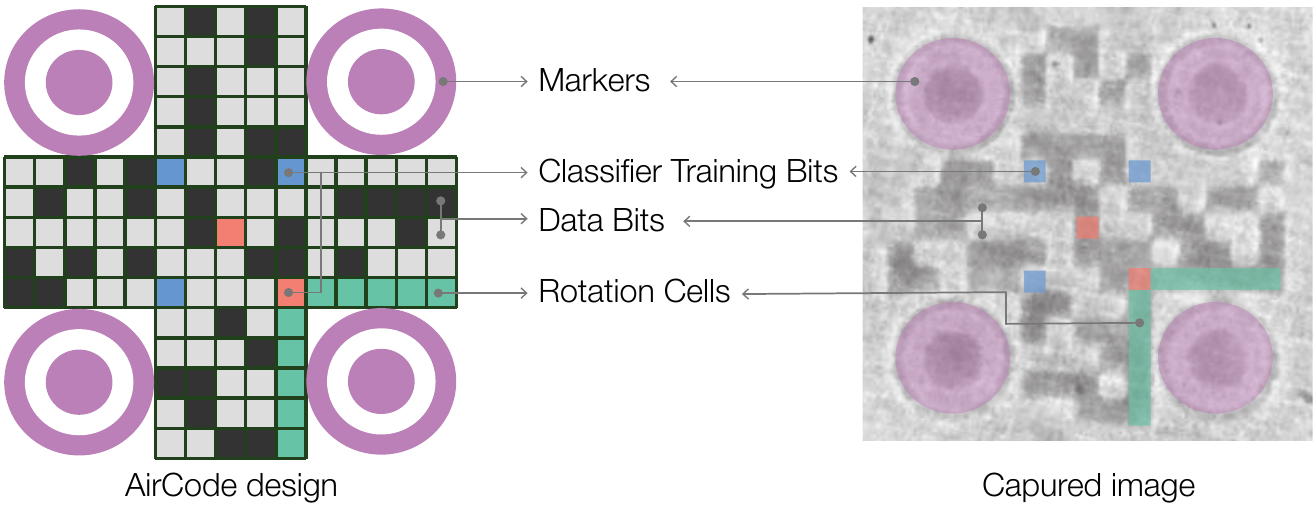}
  \caption{The structural layout of a subsurface optical code (left) and 
      the corresponding parts in a captured global-component image (right).}
  \label{fig:code_layout}
  \vspace{-4mm}
\end{figure}

\paragraph{Marker Design}
Our initial attempt was to use air pockets to assemble precisely a QR code pattern. 
But the imaging artifacts render the captured pattern unrecognizable using the standard 
QR-code decoding algorithm (\figref{fig:qrcorner}).
Instead, we adopt the concepts of markers and bits in our subsurface code
design while seeking new structures of air pockets in order to suit the
global-component decoding.
For markers, we choose to use air pockets with a concentric circular shape (\figref{fig:code_layout}),
motivated by a few observations:
Unlike QR code markers whose detection relies on the transition between black and white pixels
on a relatively clean image, we need to detect markers on a blurry and noisy image. 
The concentric circular air pockets offer unique features that can be easily
detected from the global-component image;
regardless of the blurriness, a circle always appears to be circular. 
We will present a reliable marker detection algorithm later in this section.


\paragraph{Code Generation}
Next, we generate an air pocket structure incorporating markers and bits.
Air pockets are organized in a square subsurface region, 
whose geometric parameters are estimated later in this paper.
The layout of the subsurface squares is shown in \figref{fig:code_layout}.
Four concentric circular markers are placed in the purple areas.
The center locations of these markers set up a grid of cells. In each cell, we
place a square air pocket to represent 1 or fill with solid printing material to represent 0.
Moreover, in a captured image, the entire square region may be rotated. In order to eliminate the rotation,
we identify the bottom-right marker by placing air pockets in the cells around it 
(the green cells in \figref{fig:code_layout}).
More remarkably, we place a few \emph{known} bits. The blue cells are always filled with printing material 
as bits of 1, while the orange cells are filled with air pockets to indicate bits of 0.
These bits are scattered on the grid to enable on-the-fly supervised training
for our decoding algorithm (see the Decoding step later).

The remaining cells are bits carrying user-specified information. In practice, we
use an error-correction scheme (such as the Reed-Solomon coding)~\cite{lin2004error}
to add redundancy in the provided bit string to further improve the robustness of the decoding step.
We also note that the grid resolution is user adjustable. A higher resolution accommodates 
more bits but is more susceptible to noise. 
In the Results section, we will experimentally validate different resolutions.

In our examples, the physical size of an \papername~tag is around 2cm$\times$2cm, allowing the tag to accommodate 
about 106 bits. But similar to QR codes, the number of bits is extensible. 
If a larger information capacity is needed, the user can enlarge the tag to
store more bits.  For example, as shown in the supplementary document, a
5cm$\times$5cm tag stores about 500 bits.




\subsection{Marker Detection}\label{sec:marker}
At the decoding stage, we image the physical object and 
then process the global-component image to retrieve the embedded tags. 
The first step of this process is detecting markers, consisting of three stages 
(A pseudocode of this algorithm is outlined in Algorithm~1 of the supplementary document).


\textbf{Ellipse Detector} locates individual markers from the global component
image. The global component image may have non-uniform intensity
and noise due to imaging and printing artifacts. 
We first remove intensity variations by subtracting a two-dimensional
quadratic polynomial fitted to the captured object.
We also observe that the concentric circular air pockets
become blurred and possibly distorted (if the code is not frontally facing the camera). Therefore, we adopt
the ellipse detector~\cite{OuelletH09} to detect ellipses on the image.
Exploiting the dual conic representation of an ellipse,
this method estimates elliptic parameters using image gradient, sidestepping
the detection of edge points of the ellipse. 
This feature particularly suits our problem, as the marker's boundary are low-contrast and blurry due to
the subsurface scattering.

We use the ellipse detector in a multi-scale manner to improve its robustness
We build a Gaussian pyramid of the image and detect ellipses at each scale. The
detected ellipses are then filtered based on a threshold of the ratio between
the major and minor axis lengths (in practice we set the threshold as 1.8).
At the end of this stage, we group the centers of
the ellipses if they are within a distance $\tau$ (in practice $\tau\approx5$ 
pixels), and compute the averaged center
position for each group.

\textbf{Marker Pruning} further filters the remaining ellipse centers and identifies true
markers. We exploit the fact that the four markers must form the corners of a planar square.
On the image plane, however, their positions can be distorted because of 
the camera's perspective projection. 
But if four centers are indeed the markers, then there must exist a single 
perspective transformation (the inverse of the camera projection) which restores the 
four centers into square corners.

We transform this observation into a RANSAC-style (random sample consensus)
algorithm~\cite{fischler1981random}.  In an iterative process, we pick three
ellipse centers, denoted by their 2D coordinates $\bm{u}_1$, $\bm{u}_2$, and
$\bm{u}_3$.  Since we need at least four points to compute a full projective
transformation, we assume that the transform can be approximated by an affine
transform. We then compute an affine transformation that aligns three out of
the four square corners with the selected centers by solving
$$
\begin{bmatrix}
    \bm{u}_1 & \bm{u}_2 & \bm{u}_3 \\
    1 & 1 & 1
\end{bmatrix} = 
\begin{bmatrix}
    \mathsf{A} & \bm{b} \\
    0 & 1
\end{bmatrix}
\begin{bmatrix}
    \bm{v}_1 & \bm{v}_2 & \bm{v}_3 \\
    1 & 1 & 1
\end{bmatrix},
$$
where $\mathsf{A}$ is a $2\times 2$ matrix accounting for rotation and scale, 
and $\bm{b}$ is the translation. 
$\bm{v}_1$, $\bm{v}_2$, and $\bm{v}_3$ are three corners of a square 
(i.e., $\bm{v}_1=[0\;0]^T$, $\bm{v}_2=[1\;0]^T$, and $\bm{v}_3=[1\;1]^T$).
Solving this $6\times6$ system yields $\mathsf{A}$ and $\bm{b}$, which we then 
apply to the fourth corner $\bm{v}_4=[0\;1]$.
We repeat this iteration until 
there exists another ellipse center $\bm{u}_4$ within a distance threshold $\eta$
from the transformed corner $\mathsf{A}\bm{v}_4+\bm{b}$. Then, $\bm{u}_{1...4}$
are identified as the four markers.
%
%
Typically, RANSAC-style algorithms randomly select points to
fit the model and repeat for a predefined number of iterations. 
Fortunately, we have a small number of ellipses at the end of stage one (typically 8-12).
Therefore, we can afford to iterate through all 
combinations of three ellipse centers until we find marker locations. 

\textbf{Pose Estimation} is optional, only needed when one wishes to estimate the pose
of the 3D object in addition to decoding the information. Because we know \emph{a priori} where precisely
the markers are in the object, and our marker detection establishes correspondences between
markers in the object and markers on the image, 
we can estimate the object pose (including the rotation and translation) with respect to the camera
by solving the classic perspective-4-point-problem in computer vision~\cite{szeliski2010computer}.
In our results, we will demonstrate the use of this pose estimation in a
robotic grasping application.

 
\subsection{Decoding}\label{sec:decode}

With the four markers identified, we rectify the perspective distortion of the image,
establish a grid, and now recognize the bits in individual cells.
This recognition also needs to overcome the
challenges posed by filament patterns, noise, and uneven lighting. Especially 
because a grid cell is much smaller than the marker (recall \figref{fig:code_layout}), 
it is more vulnerable to be contaminated by artifacts. As a result, na\"{i}ve
binarization of each cell based on its average pixel value is prone to error.


We choose to use a supervised learning approach to classify grid cells into 0/1 bits.
As opposed to conventional supervised learning that requires prepared training data,
we train the classifier \emph{on the fly}.  In particular, we use a support vector
machine (SVM) because it is lightweight and easy to implement.
We take advantage of the known bits that are placed during the encoding stage to train the classifier.
These bits, shown as orange and blue cells in \figref{fig:code_layout}, form our on-the-fly training set $\mathcal{T}=\{ (\bm{x}_1,y_1),(\bm{x}_2,y_2),...\}$ for SVM,
where the subscript indexes the cell with a known bit, $y_i$ are the cell's 0/1 value, and $\bm{x}_i$ are the training feature vectors.
Each element ${x}_{ij}$ in the vector $\bm{x}_i$ stores the average of pixels that are $j$ pixels
away from the center of cell $i$.
In practice, the length of the vector $\bm{x}_i$ is chosen to be pixel number that
covers 7$\times$7 cells centered at $i$.
Each feature vector is normalized to adapt to local intensity changes.
In the classification phase, feature vectors for unknown bits are constructed in the same manner. 


A detail of locating the known bits for training is worth noting.
In the captured image, the cell layout may be rotated from what is shown in \figref{fig:code_layout}.
To eliminate the rotation, recall that only the bottom-right marker has all its surrounding
cells filled with air pockets. Thus, on the image, we look for a marker whose surrounding cells 
have the lowest average pixel value and use it as the bottom-right marker.


\section{Imaging Method}\label{sec:imaging}
\begin{figure}[t]
  \centering
  \includegraphics[width=0.8\columnwidth]{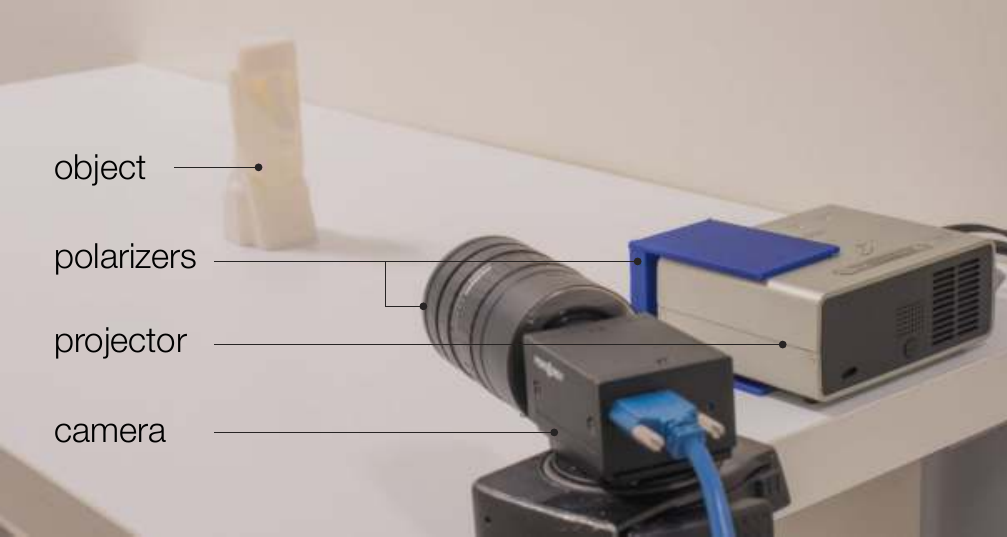}
  \caption{ {\bf Imaging setup.} 
  Our imaging system includes an off-the-shelf camera and a projector. 
  }\label{fig:setup}
  \vspace{-5mm}
\end{figure}

To read \papername~tags,
we leverage computational imaging method~\cite{NayarKGR06} to separate 
the global and direct component of light transport using structured light patterns from a projector.
This method requires only a conventional camera and projector, and the computation is fast.
Particularly, we project a checkerboard illumination pattern shifted multiple times, each producing an image.
Among the sequence of images, it computes the maximum and minimum values for each pixel, resulting
in two images $L^+$ and $L^-$.
Nayar and his coauthors showed that the direct- and global-component images can
be estimated from $L^+$ and $L^-$ using the following relationships,
\begin{equation}\label{eq:sep}
     L_d  =  L^+ - \frac{\alpha}{ 1 - \alpha}L^- ~\textrm{ and }~ L_g = \frac{L^-}{ 1 -  \alpha},
 \end{equation}
where $\alpha$ is the percentage of the activated projector pixels in the sweeping.
For checkerboard patterns that we used $\alpha = 0.5$. 
We refer to their paper for a detailed derivation of \eqref{eq:sep} 
and discuss our addition for specific needs in our problem:

\paragraph{Wavelength Choice}
Microscopically, subsurface scattering is caused by the interaction between light waves
and the material's internal irregularities (such as grain boundaries in polycrystalline solids).
As a result, the scattering behavior depends on light wavelength~\cite{bohren2008absorption},
and the longer the light wavelength is, the less likely it is scattered in a given material.

\begin{figure}[t]
  \centering
  \includegraphics[width=0.9\columnwidth]{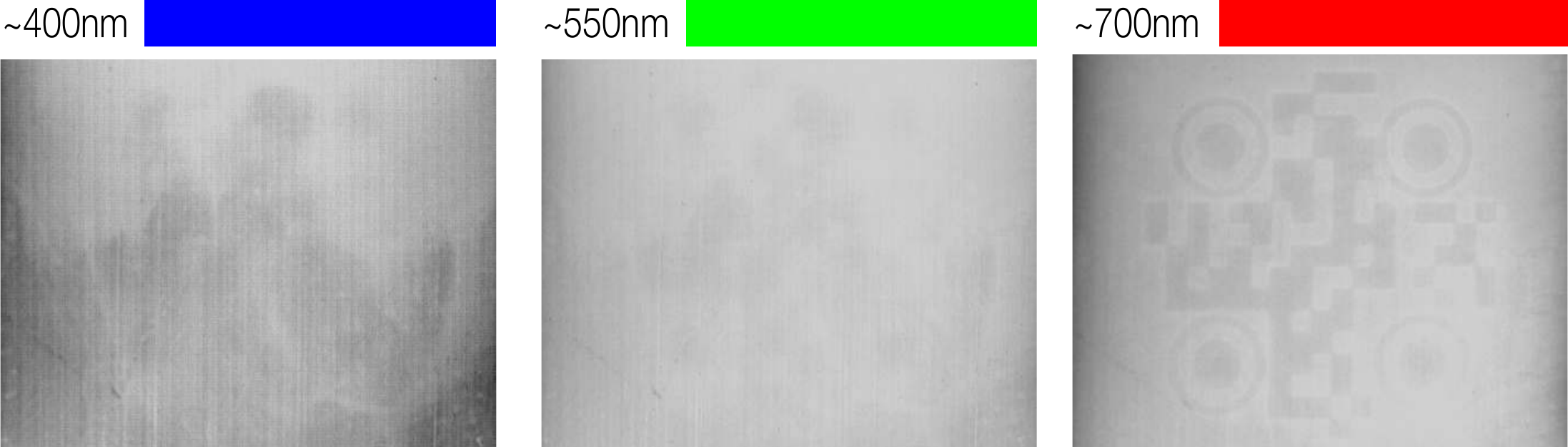}
  \vspace{-0mm}
  \caption{ 
      {\bf The effect of wavelength} on the detection of the global component. A
      longer wavelength in effect results in a longer mean free path within the
      scattering medium. As a result, the global-component image for red light
      is sharper, making the code detection and recognition easier. 
  \label{fig:wavelength}}
  \vspace{-5mm}
\end{figure}

In light of this, we take advantage of our full control of the projector and illuminate the object
with the longest light wavelength, the red light. This is because the red light is less scattered,
and penetrates deeper in the object, resulting in a less blurry global-component image.
As shown in \figref{fig:wavelength}, we compare the global component images resulted by illuminating 
with red, green, and blue light. It is evident that the red light produces an image
showing the least blurry subsurface structures.
In all of our imaging experiments, we use red light unless otherwise specified. 

\begin{figure}[b]
        \vspace{-3mm}
	\centering
	\includegraphics[width=0.95\columnwidth]{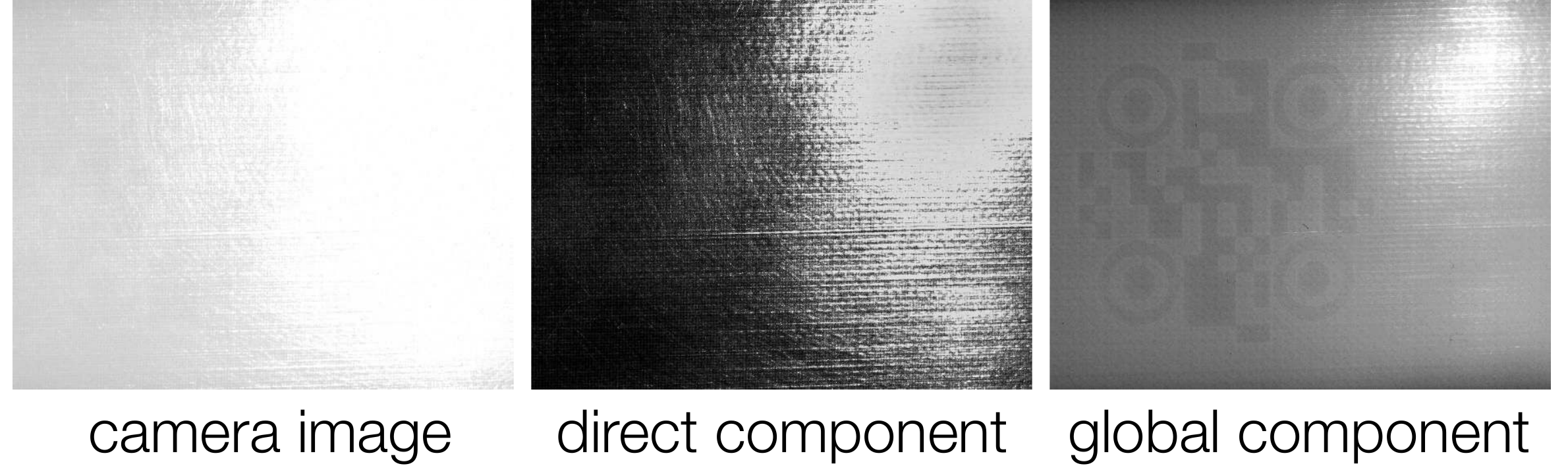}
        \vspace{0mm}
	\caption{ 
            Using the global-direct separation together with the polarizers, 
            our imaging system is robust to specular highlights.
            (left) A specular highlight is viewed in a conventional image,
            washing out most of the details on the image. 
            After separating the direct and global components, most of the 
            specular highlight remains in the direct-component image (middle).
            As a result, we can detect and recognize the \papername~tags
            from the global component image.
	\label{fig:specular}}
        \vspace{-2mm}
\end{figure}

\paragraph{Polarization}
Many 3D printed objects (in fact many objects in general) produce specular highlights
under illumination (see \figref{fig:specular}-a).
Although specular highlights are caused by reflection which mainly contributes to the direct component
of the image, in practice, they also affect the global-component image, 
because \eqnref{eq:sep}, as an estimation of the global and direct components, 
cannot separate them completely.
We mitigate the negative effect of specular light by placing linear polarizers 
in front of both the projector and the camera lens~\cite{Wolff1989,ChenLFS07}.
Because the specular reflection preserves most of its polarization but the subsurface scattering
depolarizes light, we place the two polarizers that are out of phase with each other 
to maximally eliminate the specular light. 
Combining the polarizers and the global-direct separation method, 
our imaging method is able to detect and recognize the \papername~tags, even in presence
of specular highlights (see \figref{fig:specular}).

 


\vspace{-2mm}
\section{Determining Air Pocket Parameters}\label{sec:theory}

We now present our theoretical analysis that determines air pocket parameters,
answering the questions of
\emph{how large an air pocket should be} and \emph{how deeply it should be placed
under the surface} in order to be humanly invisible but machine readable.
This analysis is a one-time process for a given fabrication material. 

To determine these parameters, we first analyze 
the change of the material's surface contrast after introducing 
subsurface air pockets
and  exploit studies of human perception of surface contrast.
Our goal here is to gain intuition on two design parameters,
namely the air pocket depth $d$ and thickness $h$ (\figref{fig:airgap}-a). 
Because human sensitivity to surface contrast is not precisely quantified, we
do not expect the analysis to give us exact design parameters but rather
identify a small range of parameters that we can choose from.

Our analysis is derived from the subsurface scattering light transport model.
In this section, we only present the key steps of this analysis 
while referring to the supplementary document for detailed derivation.


\subsection{Subsurface Scattering of Air Pockets} 
\begin{figure}[t]
  \centering
  \includegraphics[width=1.0\columnwidth]{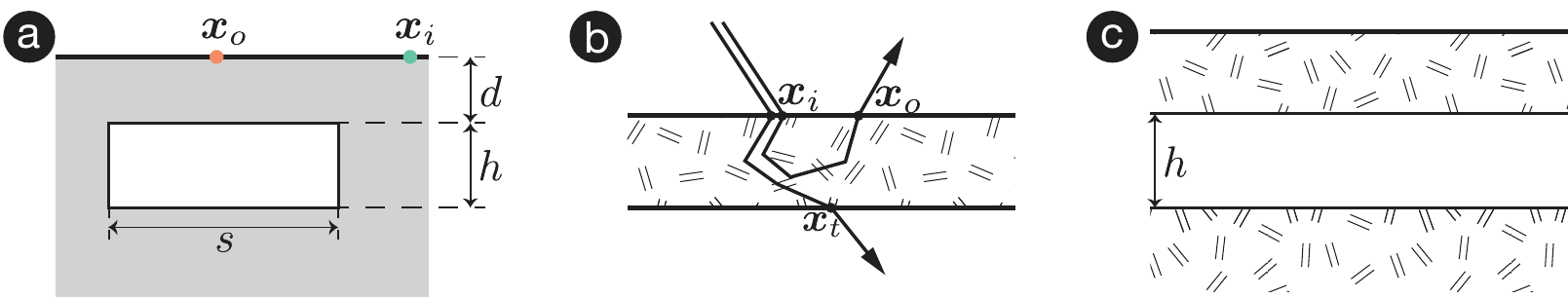}
  \vspace{-4mm}
  \caption{ Notation used for the analysis of the scattering profiles of multi-layered materials.
      \label{fig:airgap}}
  \vspace{-6mm}
\end{figure}

Consider a layer of 3D printing material. When light rays
arrive at a position of its top surface, some of them penetrate the
surface and are scattered by the substance. Eventually, some light rays pass
back out of the material from the top surface, while others escape from the
bottom surface (\figref{fig:airgap}-b).

Quantitatively, the subsurface scattering properties of a 3D printing
material~\cite{DonnerJ05} are described by its \emph{reflection} and
\emph{transmission} profiles.  The reflection profile, $R(\bm{x}_i,\bm{x}_o)$,
describes the ratio of radiant exitance (light energy) reflected by the surface
at $\bm{x}_o$ to the incident flux at $\bm{x}_i$ (\figref{fig:airgap}-b), while the
transmission profile $T(\bm{x}_i,\bm{x}_t)$ is defined similarly but for the
radiant exitance transmitted through the material layer.

Following the assumptions successfully used for modeling the subsurface
scattering of 3D printing materials~\cite{HasanFMPR10},
we consider a laterally infinite layer of homogeneous materials and almost
uniform illumination and ignore Fresnel effects.  In this case, the reflection
and transmission profiles are independent of the incident and outgoing
positions but depend only on their distance. Consequently, both
$R(\bm{x}_i,\bm{x}_o)$ and $T(\bm{x}_i,\bm{x}_t)$ can be written as 1D
functions, namely, $R(r) = R(\|\bm{x}_i-\bm{x}_o\|)$ and $T(r) = T(\|\bm{x}_i-\bm{x}_t\|)$.

Intuitively, $R(r)$ and $T(r)$ indicate how the reflected and transmitted light energy 
are distributed on the surface.
They depend on the thickness of the material layer. But we can measure 
$R(r)$ and $T(r)$ at a given thickness, following the method that has proven 
successful in~\cite{HasanFMPR10}. Then, the profile at an arbitrary thickness can be analytically 
calculated (see supplementary document for details).


%

\paragraph{Scattering Profiles of an Air Layer}
Now, consider a laterally infinite air layer of thickness $h$.
While the scattering profiles of a 3D printing material can be measured,
the scattering profiles of an air layer, to our knowledge, has not been explicitly modeled.
As noted in~\cite{jensen2001practical}, the scattering profiles 
are typically modeled by assuming light diffusive materials.
Yet, air is by no means diffusive; light travels straightforwardly in the air, not scattered at all.
In the supplementary document, we derive the transmission profile $T_\ssa(r)$ of an air layer 
and obtain
\begin{equation*}
    T_\ssa(r) = \frac{1}{A}\cdot\frac{h}{(h^2 + r^2)^{3/2}},\textrm{ where }  
    A = 2\pi\int_0^\infty \frac{h r}{(h^2 + r^2)^{3/2}}\dd r.
\end{equation*}
Here the constant $A$ normalizes $T_\ssa(r)$ 
to account for the fact that the air layer does not absorb any light. 
Note that the reflective profile $R_\ssa(r)$ of an air layer is always zero,
because an air layer never reflects light at its boundary.

\paragraph{Composition of Air Layer and Scattering Layer}
After determining the profiles of a scattering material layer and an air layer,
we can composite different layers together and compute the effective scattering
profiles.  In particular, we are interested in the scattering profiles of a
three-layer composite.  As illustrated in \figref{fig:airgap}-c, the top layer is a thin
layer of scattering material; the second layer is an air layer; the third
is a thick substrate made of scattering material also.  The effective
scattering profiles $R_\ssc(r)$ and $T_\ssc(r)$ of this composition can be
numerically evaluated by convolving profiles of individual layers.  We present the detailed derivations
and formulas in the supplementary document.


\paragraph{Scattering Profiles of Finite Air Pocket Size}
The analysis so far considers a laterally infinite air layer.
In practice, the air pocket always has a finite lateral size.
To model the reflection profile on the surface above an air pocket region (\figref{fig:airgap}-a), 
we adopt the approximation proposed by Song et al.~\shortcite{song2009subedit}
(also used by previous material fabrication work~\cite{HasanFMPR10}).
This representation defines a local profile $P_{\bx}(r)$ at a surface point $\bx$ in order to decompose the
reflection profile into $R(\bx_i,\bx_o)\approx\sqrt{P_{\bx_i}(r)P_{\bx_o}(r)}$, where 
$r = \|\bx_i-\bx_o\|$. Without air layer, the local profile
$P_{\bx}(r)$ of the homogeneous, thick material volume is $R_0(r)$.
When the laterally infinite air layer exists, the local profile 
$P_{\bx}(r)$ is the profile $R_c(r)$ for the aforementioned, triply layered material (recall \figref{fig:airgap}-c).
When the air region has a finite lateral size, the reflection profile
across the boundary of an air pocket is approximated as
$R(\bx_i,\bx_o)\approx\sqrt{ R_0(r) R_c(r)}$,
where $\bx_o$ is above the air pocket, and $\bx_i$ is in the solid material
region (\figref{fig:airgap}-a).

\subsection{Estimating Air Pocket Parameters}
\label{sec:air_pocket_analysis}
We now estimate the range of valid parameters for constructing imperceptible air pockets.
To this end, we leverage psychophysical results in visual sensory science.

\paragraph{Human Vision Sensitivity}
Particularly relevant to our design of \papername~tags are perceptual studies of 
contrast sensitivity on blurred blobs~\cite{bijl1989visibility,watt1983recognition},
because the subsurface air pockets induce a contrast of surface radiosity and the subsurface
scattering blurs their shapes.
Bijl and his coauthors~\shortcite{bijl1989visibility} discovered, through a series 
of psychophysical experiments, that (i) the human sensitivity of a Gaussian-like blob is independent 
of the blob diameter, if the view angle is larger than 20 min arc (i.e., 0.333$^\circ$) up to at least 10$^\circ$,
and that (ii) in this range the (roughly) constant contrast sensitivity threshold is 5\%.
The contrast is defined as $(L_{\textrm{max}}-L_{\textrm{min}})/L_{\textrm{min}}$, where
$L_{\textrm{max}}$ and $L_{\textrm{min}}$ are the maximum and minimum intensity of the blob, respectively.

In our problem, an \papername~tag has a size typically around 2cm.  When viewed
from a normal distance (e.g., around 75cm) away, the view angle spanned by the
codes is around 100 min arc. Exploiting the results of~\cite{bijl1989visibility}, 
we choose air pocket size and depth such that the resulting
contrast (or the change of surface radiosity) is within 5\%.
This way, by construction the air pockets are invisible to the human eye (as shown in the
Results section).

\vspace{-1.5mm}
\paragraph{Top Layer Thickness}
We now estimate the range of the top layer thickness $d$.
First, all semitransparent materials absorb light, although sometimes very weakly. 
If the top layer is too thick, most of the light rays will be absorbed before reaching the air pockets.
As $d$ increases, the influence of air pockets on the scattered light diminishes.
Given a layer of material with a thickness $d$,
the amount of light that can transmit through is quantified by its transmissive albedo $\alpha(d)$,
computed by integrating the transmission profile $T_d(r)$ over the
rotationally extruded 2D plan (i.e., $\alpha(d) = 2\pi\int_0^\infty T_d(r)r\dd r$). Here
the subscript $d$ in $T_d(r)$ is to emphasize its dependence on the thickness $d$. 
$T_d(r)$ can be computed using measured scattering profiles (see supplementary document).
The transmissive albedo $\alpha(d)$
indicates that if a top layer has a thickness $d$, then the influence of air pockets on
the surface radiosity is upper bounded by $\alpha(d)$.
Thus, we choose a $d_{\textrm{max}}$ such that $\alpha(d_{\textrm{max}})\ge\tau$, a threshold based on human
vision sensitivity ($\tau=20$\% in practice).
For the 3D printing material we use, this leads to $d_{\textrm{max}}=3$mm.

On the other end, the top layer thickness is lower bounded due to the material's mechanical strength.
If $d$ is too small, the top layer becomes fragile. In our practice, 
because  $d_{\textrm{max}}$ is already small,
we empirically test a number of $d$ (s.t. $d<d_{\textrm{max}}$) and set $d_{\textrm{min}}=1$mm.
We also note that many stress analysis methods exist to help assess the structural strength
of a 3D\rev{printed model (e.g., see~\cite{Chen:2016:stress}).}

\vspace{-1.5mm}
\paragraph{Air Pocket Thickness}
To estimate the air pocket thickness $h$,
we assume almost uniform incoming light. Then the radiosity (intensity of
outgoing light) at a surface point $\bm{x}_o$ is proportional to the surface integral
$c(\bm{x}_o)=\int_A R(\bm{x}_i,\bm{x}_o)\dd \bm{x}_i$, where $R(\bm{x}_i,\bm{x}_o)$
is the reflection profile. $R(\bm{x}_i,\bm{x}_o)$ drops quickly as the distance $\|\bm{x}_i-\bm{x}_o\|_2$ increases, 
allowing us to approximate $c(\bm{x}_o)$ by integrating over a locally flat region.

Suppose that an air pocket of thickness $h$ and lateral size $s$ is placed at distance $d$ (see \figref{fig:airgap}-a).
At the surface point $\bm{x}_o$ above the air pocket, we estimate its radiosity $c(\bm{x}_o)$ using the reflection profile 
for finite air pocket size. On the other hand, when there is no air pocket, the surface radiosity $c_0$ is approximately 
the reflective albedo of the solid thick material. 
According to the human vision sensitivity
studies~\cite{bijl1989visibility}, we define
the contrast as $(c(\bm{x}_o)-c_0)/c_0$, whose value needs to be within the
human vision contrast threshold ($\approx5$\%) to create imperceptible air pockets.
In practice, we use the upper bound, $(c(\bm{x}_o)-c_0)/c_0=0.05$, to estimate the air pocket thickness, because a larger
contrast eases machine detection of the codes. See supplementary document for detailed derivation and illustration.
In the next section, we verify that this
estimation indeed produces invisible yet machine readable tags.

\begin{figure}[t]
  \centering
  \includegraphics[width=0.95\columnwidth]{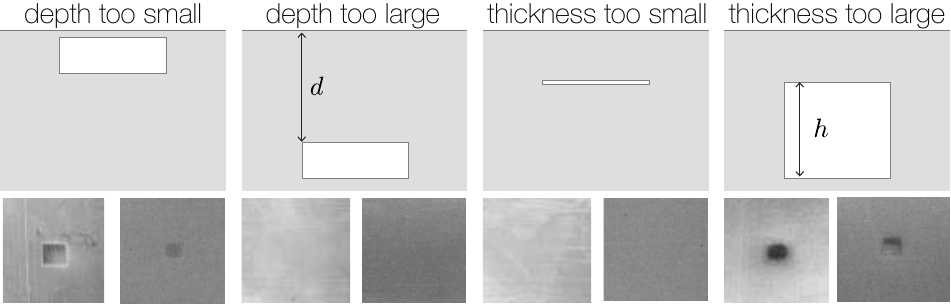}
  \caption{
      Arbitrarily choosing depth $d$ and thickness $h$  leads to unsatisfactory results.
  In each of the four pieces shown here, below the illustration, left side is the
  contrast-enhanced global-component image; right side is the photo under 
  regular lighting.
  When the depth is too small or the thickness is too large, we can 
  detect the air pockets, but they are also visible to our eyes under normal lighting.
  When depth is too large or the thickness is too small,
  the signal is too weak to be detected. 
  \label{fig:air_thinthick}}
  \vspace{-4mm}
\end{figure}

\section{Results and Validation}\label{sec:ret}

We now present the experiments we conducted to test and validate our \papername~tagging system.
We refer to the supplemental video for demonstrating the use of \papername~tags.

\paragraph{Fabrication and Imaging Setup}
We fabricated objects with subsurface air pocket tags using Stratasys Eden260VS, a PolyJet 3D printer 
with 16-micron layer accuracy (Z direction) and 200-micron planar accuracy (XY plane). 
We use a white opaque material (VeroWhitePlus, RGD835) 
and a water-soluble support 
material (SUP707).\rev{This printer can not print voids directly. 
So we printed the top and bottom
layer separately and washed away the support. 
Cylindrical connectors are added on both sides for assembly 
(see \figref{fig:show_drawer} second column).} 
The reflection and transmission profiles ($R(r)$ and $T(r)$) are measured using a 3D printed piece of
2mm thickness, following the method~\cite{HasanFMPR10}.\rev{%
Note that 
while we use a white opaque material,
our analysis is applicable to other homogeneous materials, 
such as the printing materials of other colors.
It is also possible to replace air pockets with other printing materials,
and the same analysis for air pocket parameters still applies.}

For global component imaging,
we project checkerboard illumination patterns using 
a Mitsubishi PK20 DLP projector (800$\times$600 resolution). 
Images were captured using a Point Grey Grasshopper3 monochrome linear
camera (2048$\times$1536 resolution).\rev{We use a monochrome camera 
to avoid Bayer demosaicing, as we 
consider scattering at a single wavelength.}
\figref{fig:setup} shows our imaging setup. Under low-light, the output
from this camera sensor is noisy. We, therefore, averaged 16 images for each
projected checkerboard pattern.  

\subsection{Validation}\label{sec:eval}
We validate our theoretical estimation of air pocket parameters.  
First, we fabricate pieces with unoptimized air pocket thicknesses and depths. 
As shown in \figref{fig:air_thinthick}, these arbitrarily chosen parameters lead to
either weak signals for imaging system or visible air pockets to our eyes.
Our estimation of the parameters is to ensure
that the imaging system can detect the air pockets while they remain invisible to naked eyes.
In Figure~\ref{fig:lighting} of the supplemental document, we show that the tags using our 
estimated air pocket parameters remain invisible, even under different lighting conditions and angles.

\begin{figure}[b]
    \vspace{-6mm}
    \centering
    \includegraphics[width=0.85\columnwidth]{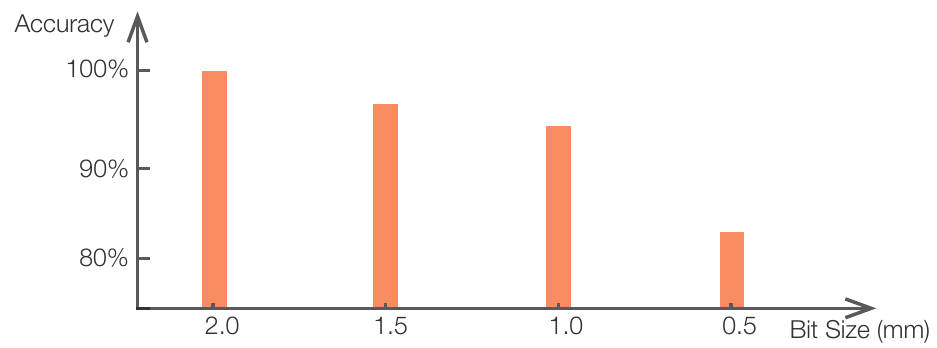}
    \vspace{-1mm}
    \caption{ 
    As the bit size (the width of each air pocket) reduces from
2mm to 0.5mm. As expected, the accuracy of the bit classification decreases with the
bit size. We stopped at  0.5mm before the quality is approaching the printer's limit.
	} \label{fig:bit_size}
\end{figure}

Next, we fabricated four testing pieces with different air pocket widths,
0.5mm, 1mm, 1.5mm, and 2mm, but the same thickness and depth.  
As shown in the Determining Air Pocket Parameters
section, a smaller air pocket width causes fewer changes of surface radiosity.
The reduced cell size, on the one hand, allows to accommodate more bits and
thereby encode more information.  On the other hand, the
decreasing contrast in the global component renders the decoding process more
difficult.  We found that the balance between increasing cell resolution and
decreasing decoding robustness is when the cell size is about 1mm (\figref{fig:bit_size}). 
For demonstrating the use of our surface codes, we choose the cell size from 1.5mm
to 2mm.  In practice, we also use Reed-Solomon code to add 40\% redundancy in
the encoded bits.

\begin{figure}[t]
	\centering
	\includegraphics[width=0.85\columnwidth]{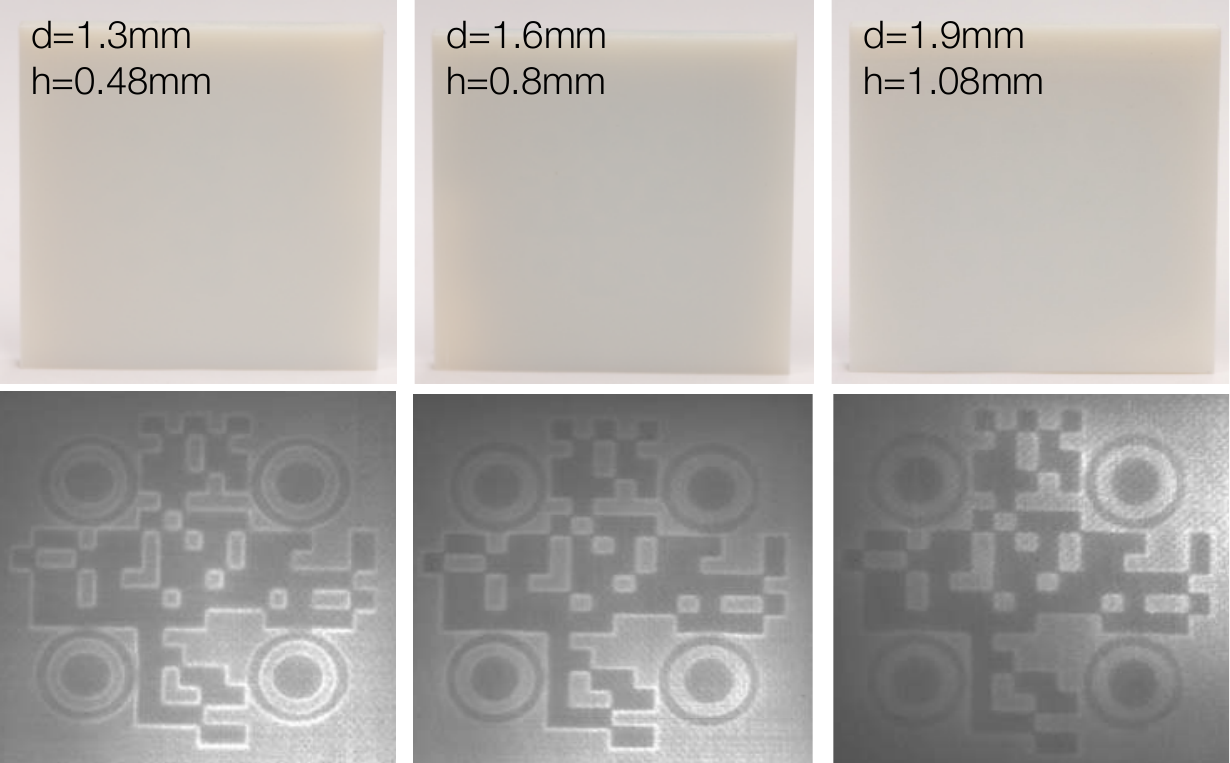}
	\caption{ 
            {\bf The thickness of the top material layer} increases from left to
            right. For each thickness, the optimal thickness of the air pocket
            computed using our model is used. In all cases, the model ensures
            that the contrast produced by the code is within the imperceptible
            range for the human eye. 
	\label{fig:depth_different_top}}
    \vspace{-5mm}
\end{figure}

\begin{figure}[t]
	\centering
	\includegraphics[width=0.999\columnwidth]{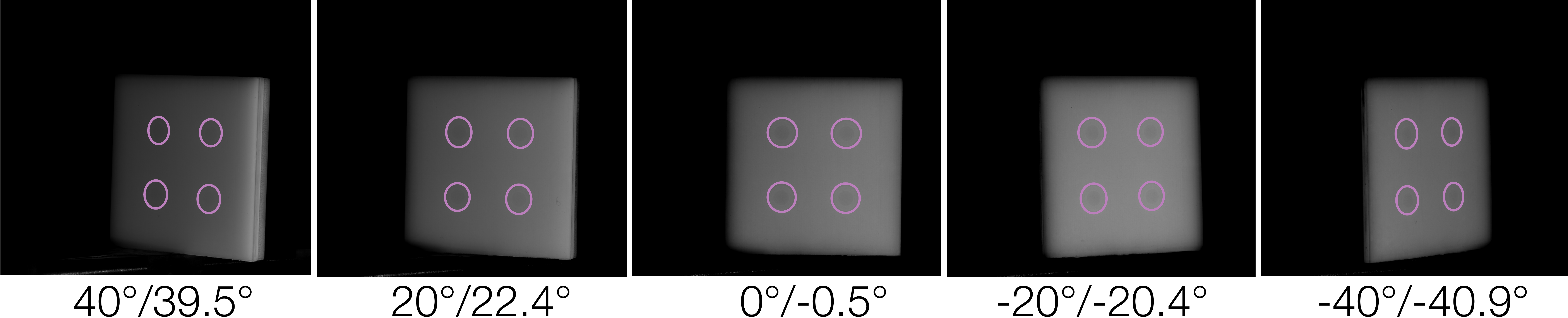}
        \vspace{-4mm}
	\caption{ 
The code detection algorithm works over a wide range of camera viewing angle
(or object pose variation). Here we see the code being successfully detected
over a +/-40 degree range of code orientation with respect to the camera. }
	\label{fig:angles}
        \vspace{-6mm}
\end{figure}



Furthermore,
we choose different top layer thickness values and estimate corresponding air layer thickness $h$ such that the 
surface radiosity contrast is about 5\% according to our model. The results are shown in~\figref{fig:depth_different_top}.
In all pieces, the subsurface codes are indeed invisible, while their global-component images all reveal
the tags and are machine readable.


To validate the robustness on orientation, 
we placed a fabricated piece on a rotary plate and tested the imaging and decoding algorithm
with different object orientations.
\figref{fig:angles} shows that our method is able to read the subsurface codes for 
rotations in the range of $[-40^{\circ},40^{\circ}]$ with respect to the camera view direction.
We note that this orientation range is much larger than many radiographic techniques.
For example, as pointed out in~\cite{WillisW13}, the Terahertz imaging system can only image objects
within +/-11$^\circ$.
Moreover, we measured the accuracy of the pose estimation using subsurface markers
(recall the last stage in the Marker Detection section).
Our result shows that we can estimate the object rotation angle with an error less than  $2^{\circ}$.

\section{Discussion and Future Work}
We have presented a method that tags physical objects with user-specified
information in a digital manufacturing process. 
These unobtrusive tags are imperceptible to the human eye but recognizable by a computer.
This is achieved by placing carefully designed air pocket structures underneath the surface of
3D printed objects. 

Although we showed that the global component image is robust against various factors,
the capturing process still takes \rev{3-4} minutes.\rev{This is because a checkerboard 
sweeping takes about 10 seconds. 
To reduce imaging noise, we take multiple sweeps and average them.} %
The imaging time can be shortened by using cameras with less sensor noise or using infrared 
projectors and cameras. With the constantly improving imaging quality of commodity cameras,
we expect that the \papername~tag reading time will be shortened in the near future.
Furthermore, we found that by projecting a red image with
cross-polarization, we can effectively remove a large number of surface
artifacts and specular highlights.  However, to recover the global component,
we still have to sweep the checkerboard pattern multiple times. It would be an
interesting future work to reduce the amount of image captures needed to read
the AirCode tags.

Our analysis assumes that the 3D printing material is largely homogeneous and semitransparent. 
While this assumption is valid for many 3D printing systems that fabricate with plastic materials, other printers cannot produce nearly homogeneous 
materials. For example, many fused-deposition modeling (FDM) printers deposit relatively thick filaments,
and the printed object is not homogeneous.\rev{Moreover, it is a common postprocess to 
paint the surface of a 3D printed object. While our method can
account for semitransparent paint, it will fail if the paint is completely
opaque. Another limitation is that,
unlike traditional optical codes that can be easily replaced if needed,
AirCode tags can not be updated after the object is printed, as they are baked into the
geometry.  But sometimes this is a desired feature to prevent the tags from
tampering.}



We can place \papername~tags under gently curved surfaces (e.g., the mug grasping in \figref{fig:robot_mug})
But objects with highly curved shapes present challenges with
respect to the air pockets design. In those cases, placing air
pockets at a fixed height below the surface with fixed depths will not always
produce recognizable tags. 
We intend to explore other computational methods to compute complicated
depth fields that can generate desired global-component effects.



\hypersetup{%
  pdftitle={\plaintitle},
  pdfauthor={\emptyauthor},
  pdfkeywords={\plainkeywords},
  pdfdisplaydoctitle=true, 
  bookmarksnumbered,
  pdfstartview={FitH},
  colorlinks,
  citecolor=black,
  filecolor=black,
  linkcolor=black,
  urlcolor=black,
  breaklinks=true,
  hypertexnames=false
}

\section{Acknowledgements}

We thank the anonymous reviewers for their feedback.
We are grateful to Arthur Autz for the support of 3D printing facilities,
Daniel Miau and Brian A. Smith for the feedback on writing,
Henrique Teles Maia for proofreading and narration,
Yonghao Yue and Shuang Zhao for rendering suggestions,
Klint Qinami and Anne Fleming for proofreading an early draft,
Daniel Sims for managing hardware equipment,
and Jason Hollaway for imaging and hardware suggestions.
This work was supported in part by the NSF Award CAREER-1453101. 
Dingzeyu Li was partially supported by an Adobe Research Fellowship.

\balance{}

\bibliographystyle{SIGCHI-Reference-Format}
\bibliography{ref}

\newpage

{\Large Supplemental Document:}

\section{Background in Subsurface Scattering}

When light rays arrive at the material surface, some of them penetrate the surface and 
are scattered by the substance, before passing back out of the material. 
The subsurface scattering has been widely modeled
by the bidirectional subsurface scattering reflectance distribution function (BSSRDF) 
an eight-dimensional function, denoted as $S(\bm{x}_i,\bm{\omega}_i,\bx_o,\bm{\omega}_o)$,
that determines the outgoing radiance $L(\bm{x}_o,\bm{\omega}_o)$ at a surface point $\bx_o$
in direction $\bm{\omega}_o$ using
$$
L(\bm{x}_o,\bm{\omega}_o) = \int_A\int_{\Omega^+}L(\bx_i,\bomg)S(\bx_i,\bomg_i,\bx_o,\bomg_o)\cos\theta_i \dd\bomg_i\dd\bx_i,
$$
where $L(\bx_i,\bomg_i)$ is the incoming radiance arriving at a surface point $\bx_i$ from 
direction $\bomg_i$, and $\theta_i$ is the angle between the surface normal direction and $\bomg_i$.
Following the assumptions successfully used in the fabrication of subsurface
scattering materials~\cite{HasanFMPR10},
we consider homogeneous materials and almost uniform illumination, and ignore Fresnel effects. 
Then, the BSSRDF reduces to
a 1D radial function, $S(r) = S(\|\bx_o-\bx_i\|)$.
Moreover, we adopt the notions of \emph{reflection} and
\emph{transmission} profiles~\cite{DonnerJ05}, denoted as $R(r)$ and $T(r)$, respectively.
$R(r)$ is the ratio of the radiant exitance reflected by the surface to the incident flux,
while $T(r)$ is defined similarly but for the radiant exitance transmitted
through the material volume. 
$R(r)$ and $T(r)$ are related to $S(r)$ through $R(r) = \pi S^+(r)$ and $T(r) = \pi S^-(r)$,
where ``+'' and ``-'' indicate whether the incoming and outgoing point are on the same
or opposite side of the material volume.

With these notions, we can express the reflection and transmission profile of a
multi-layered material using the profiles of its individual layer components. 
Consider two laterally infinite slabs of constant 
thickness (see \figref{fig:airgap}-b). 
Let $R_i$ and $T_i$ ($i=1,2$) denote the profiles of these two slabs. 
Attaching the second slab to the bottom side of the first one
results in a new slab with the scattering profiles related to
$R_i$ and $T_i$ through convolutions~\cite{DonnerJ05}. In short, the scattering
profiles of the new slab are 
\begin{equation}\label{eq:composite}
\calR_{12} = \calR_1 + \frac{\calT_1\calR_1\calT_1}{1 - \calR_1\calR_2}
\;\textrm{ and }\;
\calT_{12} = \frac{\calT_1\calT_2}{1-\calT_1\calT_2},
\end{equation}
where $\calR$ and $\calT$ are the Hankel transform---the equivalent of Fourier transform 
for radially symmetric functions---of $R$ and $T$, respectively~\cite{HasanFMPR10}.
When there is not confusion, we will refer $\calR$ and $\calT$ also as the reflection and transmission
profiles. 




\section{Computing Transmission Profile $T_d(r)$}\label{app:trans}
We follow the Kubelka-Munk theory to compute the transmission profile $T_d(r)$
of a laterally infinite slab with a thickness $d$~\cite{cortat2004kubelka,HasanFMPR10}.
We measure the reflection and transmission profiles $R_D(r)$ and $T_D(r)$ of a slab with a thickness $D$,
and compute their Hankel transforms $\calR_D$ and $\calT_D$. Then, $\calR_d$ and $\calT_d$ of
an arbitrary thickness $d$ can be computed as
$$
\calR_d = \frac{S(d)}{\alpha S(d) + \beta C(d)}\;\text{ and }\; 
\calT_d = \frac{\beta}{\alpha S(d) + \beta C(d)},
$$
where $S(d)=\sinh(\gamma d)$ and $C=\cosh(\gamma d)$ are hyperbolic functions,
and 
$$
\alpha = 1 + \frac{\mathcal{K}}{\mathcal{C}},\;\;\beta=\sqrt{\alpha^2-1},\;\;\gamma=\sqrt{\mathcal{K}(\mathcal{K}+2\mathcal{S})}.
$$
Here both $\mathcal{S}$ and $\mathcal{K}$ are constant values, computed by
$$
\mathcal{S} = \lim_{d\to0}\frac{\mathcal{R}_d}{d}\;\;\text{ and }\;\;
\mathcal{K} = \lim_{d\to0}\frac{1-\mathcal{R}_d-\mathcal{T}_d}{d}.
$$
The limits here can be computed by repeatedly halving the thickness $d$ (starting from $\mathcal{R}_D$ and $\mathcal{T}_D$)
using the relationship~\eqref{eq:composite}.

\section{Pseudocode for Marker Detection}

\begin{algorithm}[H]
\caption{Marker Detection}
\label{alg:marker}
\begin{algorithmic}[1]
\Procedure{Marker Detection}{}
\State Pre-Process the image
\For{each scale in scale\_range}
        \State Detect ellipses
        \State Append ellipse centers to candidate pool
\EndFor
\State Group duplicates from the candidate pool
\For{every combination of 3 ellipses$(nC_3)$} 
\State Compute affine transform
\State Apply transform to $4^{th}$ corner marker
\State Search candidate pool for match
\If {match found}
\State Compile detected marker centers
\State break;
\EndIf
\EndFor
\If {valid markers found}
\State Compute projective transform using all 4 markers
\State Rectify image using the transform
\EndIf
\State Estimate object pose 
\EndProcedure
\end{algorithmic}
\end{algorithm}

\begin{figure}[b]
  \centering
  \includegraphics[width=0.99\columnwidth]{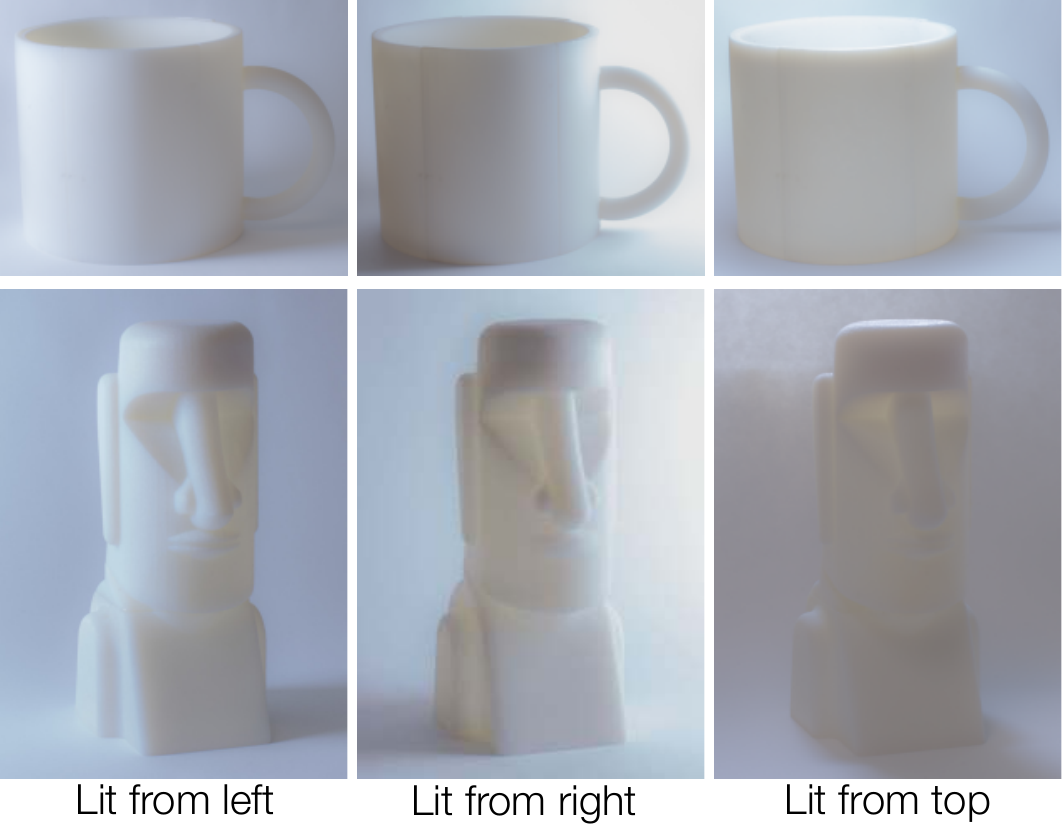}
  \caption{ 
      Under different extreme lighting, the embedded AirCode tags remain invisible.
      We capture these photos while moving a strong light source around
      the statue and the mug. The light source we use is a 45W photography light bulb at 6500K.
      \label{fig:lighting}}
\end{figure}

\section{More Experimental Analysis}

\figref{fig:lighting} demonstrates the proposed AirCode remains imperceptible under extreme lighting
conditions.

\begin{figure}[t]
  \centering
  \includegraphics[width=0.99\columnwidth]{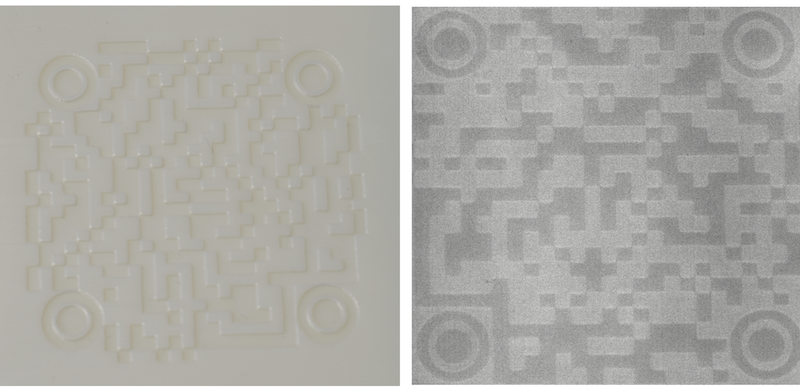}
  \caption{ 
      AirCode with larger capacity. On the left is fabricated piece and on the right is the 
      imaging global-component image.
      \label{fig:high_cap}}
\end{figure}

As mentioned in the main text, AirCode naturally extends to handle higher data capacity, similar
to QR code design. In \figref{fig:high_cap}, a larger tag is designed and printed. More than 500 bits
are encoded in this 5cm by 5cm AirCode tag.

\begin{figure}[t]
	\centering
	\includegraphics[width=0.99\columnwidth]{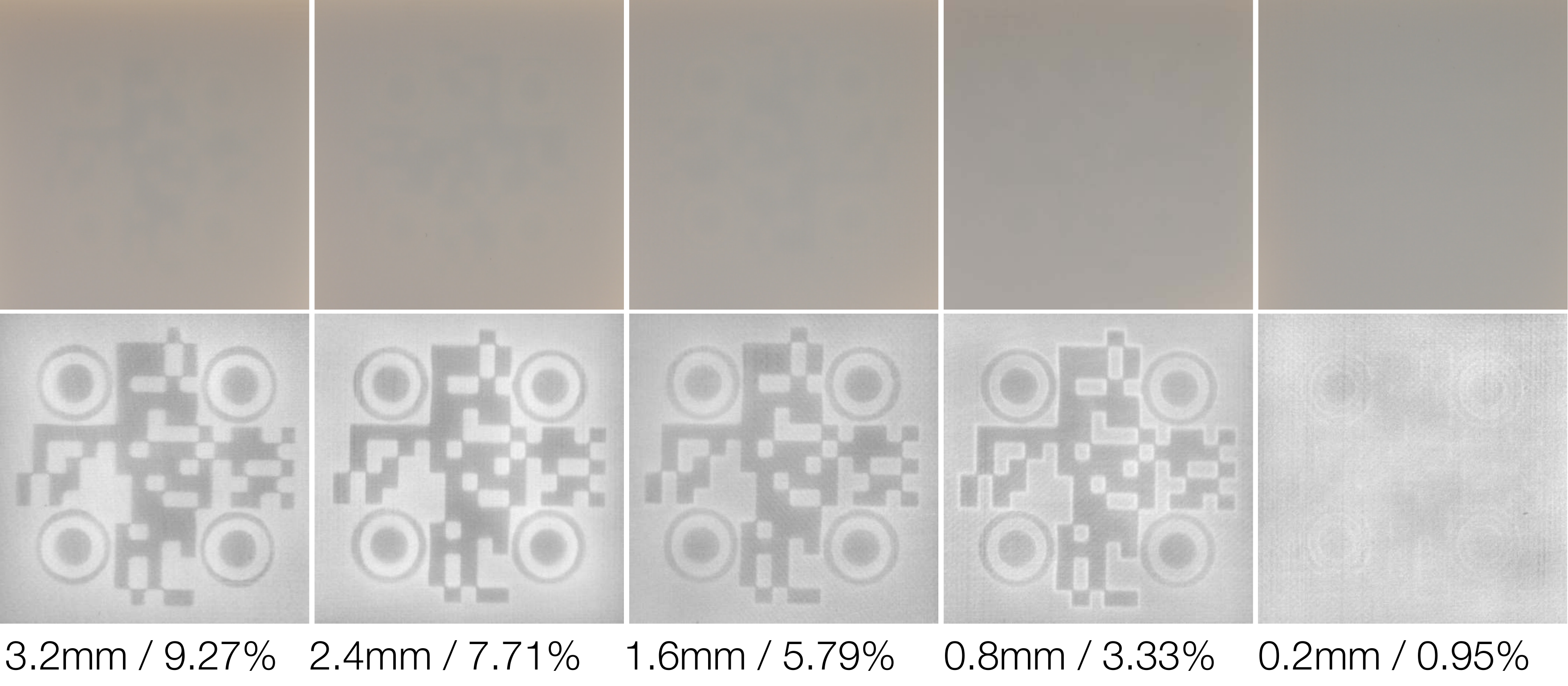}
        \vspace{-1mm}
	\caption{ 
            {\bf Visibility of embedded codes} as a function of the thickness of the
            air pocket. The thickness of the air pocket decreases from left to
            right, while the thickness of the top material layer remains unchanged.
            The number on the left of each column indicates the air pocket thickness,
            while the number of the right indicates surface radiosity contrast computed
            using our model.
        \label{fig:depth_same_top}}
        \vspace{-1mm}
\end{figure}

To analyze the visibility quantitatively, 
we fabricate pieces with different air pocket thicknesses and depths. 
To show As the air thickness decreases, the air layer transmission profile $T_\ssa(r)$
Consequently, the air pockets become less visible, as shown in~\figref{fig:depth_same_top}.
If the air pocket is too thin (0.2mm), the resulting contrast cannot be detected
by the algorithm, as the signal-to-noise ratio is too low.

\begin{figure}[t]
	\centering
	\includegraphics[width=0.99\columnwidth]{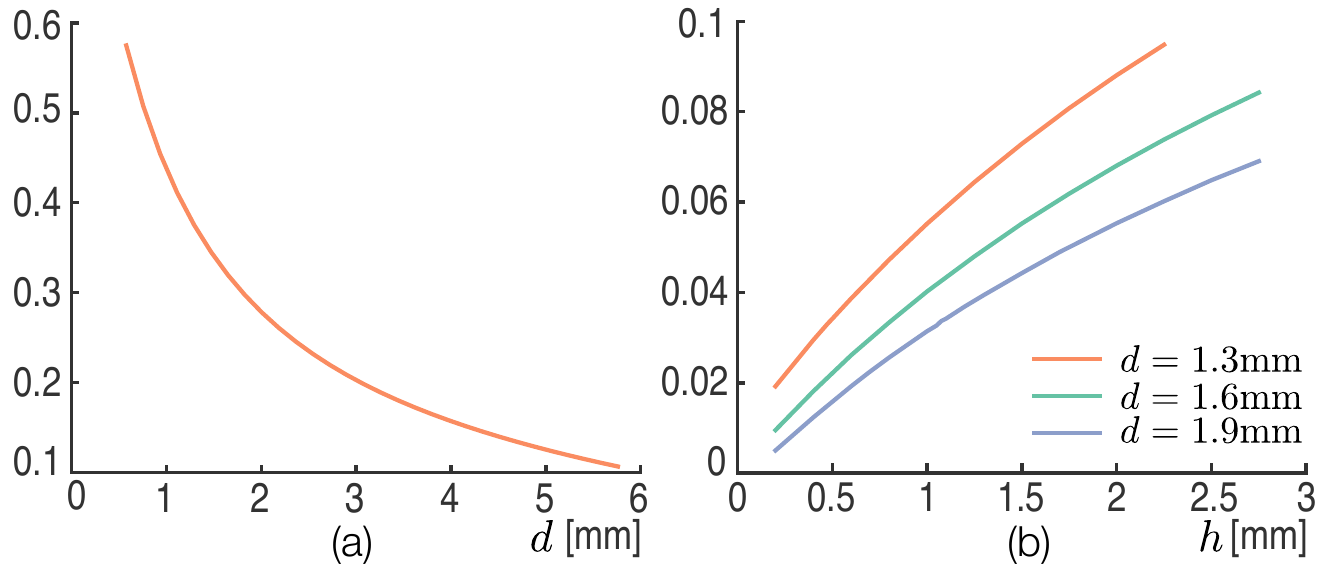}
        \vspace{-1mm}
	\caption{ 
            (a) The transmissive albedo of a scattering material (the 3D printing material we used) as a function of its thickness d. (b) The contrast between a surface point above an air pocket with respect to a surface point without one, plotted as a function of the thickness h of the air pocket. The three curves are plotted for differ- ent thickness d of top material layer.
        \label{fig:curve}}
        \vspace{-1mm}
\end{figure}

The transmissive albedo $\alpha(d)$ provides us
insight on setting the maximum of $d$: it indicates that if the top layer
has a thickness $d$, then the influence of air pockets on the surface
radiosity is upper bounded by $\alpha(d)$. Thus, we choose a $d_{max}$ such
that $\alpha(d_{max}) \geq \tau $, a threshold based on human vision sensitivity
( $\tau  = 20\%$ in practice). For example, for the 3D printing material
 we use, this leads to $d_{max} = 3mm$.
In \figref{fig:curve}-b, we consider an air pocket with a lateral size $s = 4mm$
 and visualize the change of the contrast value with respect to $h$ and $d$.

\end{document}